\documentclass[12pt]{article}
\usepackage{times}
\usepackage{graphicx}
\usepackage[percent]{overpic}
\usepackage{amssymb}
\usepackage{import}
\usepackage{amsmath,bm}
\usepackage{lineno}
\usepackage{subcaption}
\usepackage{tikz}
\usepackage{subfiles}
\usepackage{soul}
\usepackage{xcolor}
\usepackage{pgfplots}
\usepackage{siunitx}
\usepackage[labelfont=bf]{caption}

\usetikzlibrary{spy, calc}
\pgfplotsset{width=7cm,compat=1.14}

\usepackage{zb-se}

\title{Spike-Triggered Descent}
\author
{Michael Kummer and Arunava Banerjee\\
\\
Computer and Information Science and Engineering, University of Florida,\\
\\
PO Box 116120, Gainesville, FL 32611, USA\\
\\
E-mail: msk@cise.ufl.edu, arunava@ufl.edu
}
\date{}

\def\biblio{\bibliographystyle{agsm}\bibliography{./refs}}
\graphicspath{{./img/}}

\begin{document}
\def\biblio{}

\baselineskip24pt

\maketitle

\begin{abstract}
\normalsize
\baselineskip24pt
The characterization of neural responses to sensory stimuli is a central problem in neuroscience.  Spike-triggered average (STA), an influential technique, has been used to extract optimal linear kernels in a variety of animal subjects.  However, when the model assumptions are not met, it can lead to misleading and imprecise results.  We introduce a technique, called spike-triggered descent (STD), which can be used alone or in conjunction with STA to increase precision and yield success in scenarios where STA fails.  STD works by simulating a model neuron that learns to reproduce the observed spike train.  Learning is achieved via parameter optimization that relies on a metric induced on the space of spike trains modeled as a novel inner product space.  This technique can precisely learn higher order kernels using limited data.  Kernels extracted from a \textit{Locusta migratoria} tympanal nerve dataset\cite{https://doi.org/10.6080/k0bg2kwb} demonstrate the strength of this approach.
\end{abstract}

\biblio

\section*{Introduction}

A major goal of sensory neuroscience is to precisely characterize the mapping that specifies how a neuron responds to sensory stimuli.  This response function accounts for intermediary physical processes along with activity of the entire network upstream from the target neuron.  The complexity of this problem arises from variations in network connectivity and constituent ion channels which cause wildly differing behavior.  There's additional difficulty in a direct component wise analysis resulting from discontinuity of the spiking behavior caused by Hodgkin Huxley ion channels \cite{hodgkin1952quantitative}.  The inaccessible and innumerable physical parameters create complex interactions which lead to intractable calculations necessitating model simplifications.

A key simplifying assumption comes from signal processing: Any time-invariant continuous nonlinear operator with fading memory can be approximated by a Volterra series operator \cite{boyd1985fading}.  The overall impact upon the membrane potential by the upstream network can then be described by a set of Volterra Weiner kernels \cite{wiener1966nonlinear} \cite{bialek1991reading} \cite{rieke1997spikes}.  Using the first order kernel: Spike-triggered average (STA), a technique which has seen widespread application, assumes a simple probabilistic model of spike generation.   When the model assumptions are met, it returns an optimal first order kernel.  We introduce a new technique, called spike-triggered descent (STD), which can learn higher order kernels and yield higher accuracy.  These techniques approximate the desired kernel by constructing a relationship between sensory stimuli and the spike trains they cause.

Here we give an overview of how STA \cite{de1968triggered} \cite{marmarelis1972white} \cite{chichilnisky2001simple} \cite{schwartz2006spike} and STD work, describe their models, and point out a few key differences.  STA is based on the linear non-linear Poisson cascade model (LNP) which convolves the signal with a linear kernel, applies a point nonlinearity to convert it into a firing rate, and then samples it using an inhomogeneous Poisson point process to generate spikes.  Obtaining the kernel from the signal and resulting spikes requires the signal to be a stationary Gaussian process so that Bussgang's theorem \cite{bussgang1952crosscorrelation} can be applied.  In contrast to STA, STD is based on the cumulative spike response model (CSRM) \cite{gerstner1996matters} and can approximate higher order kernels for any sufficiently complex signal.  Replacing the nonlinear Poisson spike generation, the CSRM spikes occur when the convolution's resulting membrane potential exceeds threshold which inhibits future spikes by way of an after hyperpolarizing potential (AHP).  STD works by comparing simulated and recorded spike trains to form a gradient that optimizes kernel parameters.

Neuroscientists use STA because it recovers the optimal linear kernel and is easy to use.  It has been applied in a variety of applications including creating bionic interfaces\cite{rathbun2018spike}.  However, the reliance on the stimulus being a Gaussian process and the restriction to first order kernels are weaknesses which STD does not share.  The techniques can be used in tandem or STD's initial learning kernel can be randomly guessed.  Surprisingly without over fitting, it can precisely approximate while reusing limited data.  In the following sections, we introduce a spike train metric used in momentum based stochastic gradient descent (SGD) to update kernels that represent response functions.

\biblio

\section*{Results}
Spike-triggered descent is a robust technique that can precisely approximate linear and higher order kernels along with an AHP time constant.  After showing a case where STD excels beyond STA, we indicate why.  Introduced next is a generalization of the cumulative spike response model (GCSRM).  At the core of STD is the ability to minimize the distance \eqref{eq:distance} between the desired ($D$, experimentally measured) and output ($O$, simulated reconstruction) spike trains with respect to kernel parameters.  As the simulated neuron's spike trains approach the desired, so too do its parameters.  Convergence is demonstrated in a variety of experimental setups: first and second order kernels, first order kernels with accompanied AHP time constants, and first order kernels from LNP spikes.  Additionally demonstrated is the robustness against noise effects applied to the input, spike times, and spike addition/deletion.  Finally, we demonstrate the effectiveness by applying it to a \textit{Locusta migratora} dataset.

\subsection*{STD compared to STA}

\begin{figure}
\centering
\scalebox{1.0}{
\begin{subfigure}{.175\textwidth}
\includegraphics[scale=1.0]{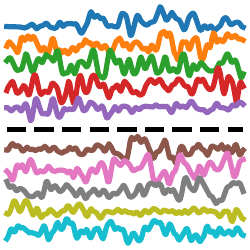}
\caption{\label{fig:std_vs_sta_a}}
\end{subfigure}%
\begin{subfigure}{.65\textwidth}
\includegraphics[scale=1.0]{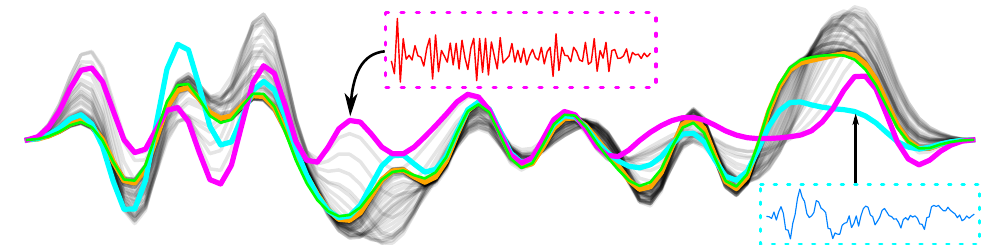}
\caption{\label{fig:std_vs_sta_b}}
\end{subfigure}%
\begin{subfigure}{.2\textwidth}
\includegraphics[scale=1.0]{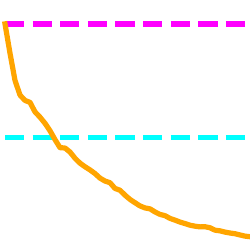}
\caption{\label{fig:std_vs_sta_c}}
\end{subfigure}%
}
 \caption{ \label{fig:U2D_STA_STD} \textbf{a}, The input is distorted by passing it through a second order kernel modeled by a grid made of $20 \times 20$ third order cardinal B-splines.  This produces a new signal that is used instead as input $s = \int \int K x(t-\tau_1) x(t-\tau_2) d\tau_1 d\tau_2$.  The first and last $5$ of $100$ input slices of $200ms$ are shown sequentially.  \textbf{b}, The distorted input leads to a high frequency STA kernel (red) which is then smoothed (magenta).  STD proceeds from here by iterating through a chain of intermediary learning kernels (black) to approximate (orange) the desired (green).  The undistorted input's result before (blue) and after (cyan) smoothing show the effect the distortion has on STA. \textbf{c}, The error, as measured by $\frac{ |k_{des} - k_{lrn}| } { |k_{des}| }$, decreases to $8\%$ (orange) as the STD kernel approaches the desired.  Smoothing and scaling the undistorted and distorted STA results improves them from $58\%$ (blue) to $50\%$ (cyan) and from $119\%$ (red) to $98\%$ (magenta). }
\end{figure}

Shown in Figure \ref{fig:U2D_STA_STD} are results from an experimental setup where STD achieves superior accuracy as compared to STA in two distinct experiments (before and after distortion).  The input signal to this virtual neuron was uniform white noise sampled at 1kHz distorted by a second order kernel.  The stimulus drove a neuron which had a first order kernel and spiked when the convolved signal crossed a fixed threshold where each spike had an impact according to an exponential decay function describing an AHP.  The STA kernel yields an error of $50\%$ (cyan) as defined by $\frac{ |k_{des} - k_{lrn}| } { |k_{des}| }$ where $|\cdot|$ is the $L_2$ norm.

Adding signal distortion via a second order kernel creates a signal (\ref{fig:std_vs_sta_a}) that results in STA having high frequency oscillation about the origin with an error of $119\%$ (red) which when smoothed had an error of $98\%$ (magenta).  For comparison purposes, instead of randomly initializing, STD is using the scaled and smoothed STA result (magenta) as a starting place for gradient updates until convergence.  This proceeds by randomly selecting one of the $100$ input slices (\ref{fig:std_vs_sta_a}) $10,000$ times and eventually yields an $8\%$ error (orange).  To reiterate, STD achieved superior results while retraining on the same data and not over fitting!  When trying to obtain a close approximation, to avoid compounding errors in complex analyses, it would be favorable to use STD.

\subsection*{Why STA requires Gaussian input}

\begin{figure}
\centering
\includegraphics[scale=0.28]{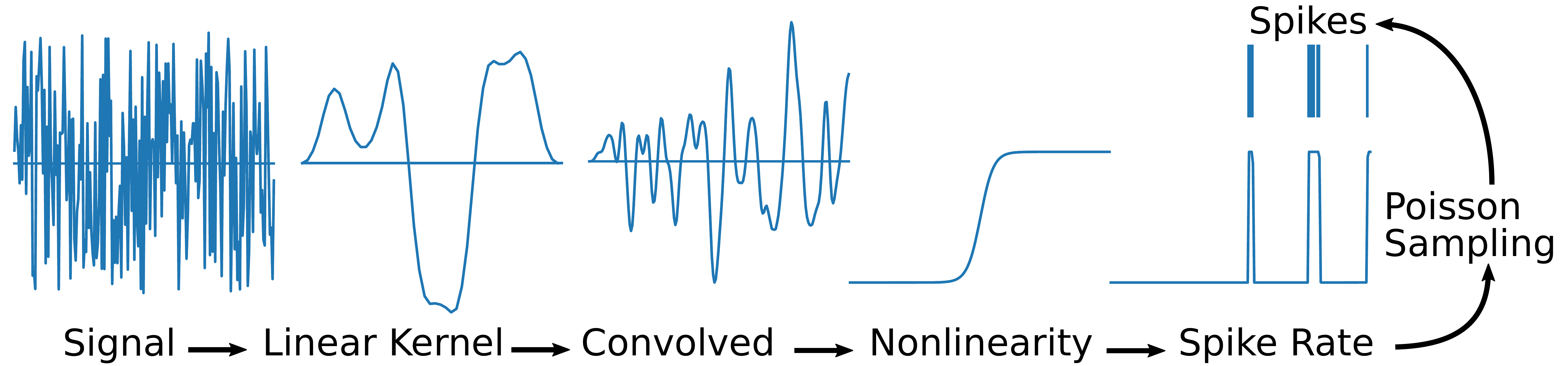}
\caption{\label{fig:LNP_spike_process} A signal is convolved with a linear kernel (L), passed through a nonlinearity (N) to achieve a spike rate, and then Poisson sampled (P) to produce spikes.}
\end{figure}

A graphical summary of how the LNP model operates is shown in Figure \ref{fig:LNP_spike_process}.  The linear component is the first order Volterra kernel $k$, the nonlinear component (N) is a function $v$ that maps the kernel response $y = k * x_w$ to a firing rate $z$, and an inhomogeneous Poisson (P) process generates spikes.  At any point in time, the probability of firing is $z = v(k * x_w)$.  The solution to the STA kernel can be expressed as $k = \phi_{xx}^{-1} \phi_{xy}$ where $\phi_{xx}$ is the signal's autocorrelation and $\phi_{xy}$ is the correlation of the signal with the potential.  Instead of $\phi_{xy}$, we can instead only observe a correlation with the spike rate $\phi_{xz}$.  Bussgang\cite{bussgang1952crosscorrelation} showed $\phi_{xy} = C \phi_{xz}$ the resulting kernel differs only by a scaling factor after the nonlinearity is applied, when the input is a Gaussian process.

\subsection*{Generalized cumulative spike response model}

\begin{figure}
\centering
\includegraphics[scale=0.28]{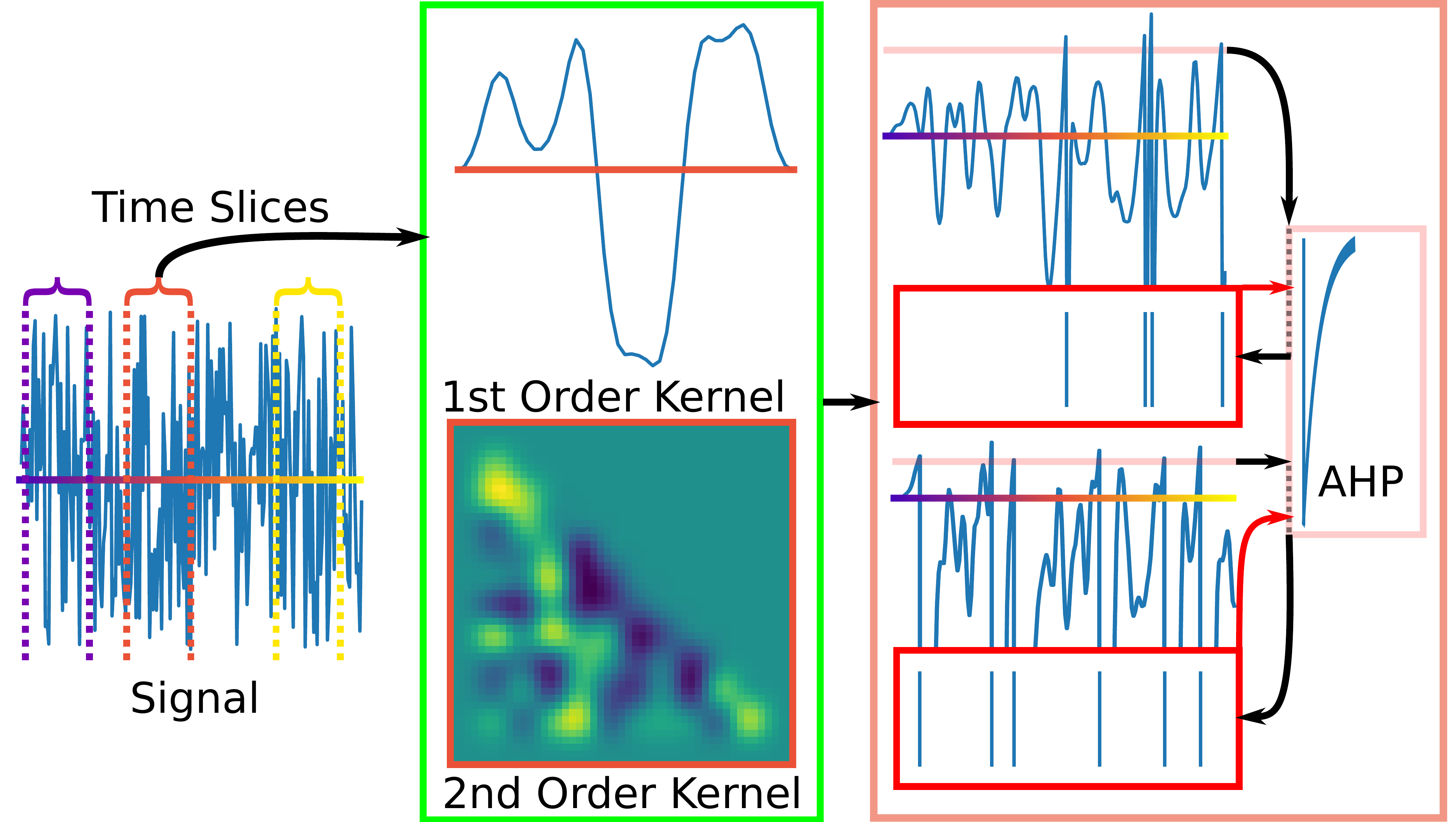}
\caption{ \label{fig:AHP_spike_process} Slices of the uniformly random generated signal are serially passed through a kernel to produce a simulated electrical potential.  When above threshold (pink) an AHP, with time constant $\mu$ generally set to $1.2ms$, is applied and a spike is produced which collectively (red box) inhibit future firing.}
\end{figure}

The cumulative spike response model (CSRM) \cite{gerstner1996matters} illustrated in Figure \ref{fig:AHP_spike_process} can be generalized to include the full Volterra series of kernels \eqref{eq:GCSRM}.  These kernels, approximated here by splines, represent the impact on the membrane potential by the stimuli's higher order auto-correlates.  Increasing $n$ allows for increased pattern detection capabilities.  The AHP function $\eta = -A e^{(t_l - t_k)/\mu}$ models the refractory period which is a region where spikes are highly unlikely to occur right after firing.  In contrast to Poisson sampling, the AHP approach is deterministic and imposes a prior state dependency.  A spike is generated when a threshold ($\tilde{\Theta}$) is exceeded by the signal convolved with kernel(s) minus past spikes' AHPs.  The n-order kernel $K_n$ is a spline function composed of the n-ary Cartesian product of third order cardinal B-splines $B_n$.  The kernel is incrementally updated by the optimization process.  At time $t=t^O_l$ the kernel is $K_{l,n} = \sum_{i} B_{i,l,n}\beta_{i,l,n}$.  The current and prior spike times are $t^O_l$ and $t^O_k$.  The AHP parameter $\mu$ modifies the refractory time.

\begin{equation} \label{eq:GCSRM}
\begin{split}
\tilde{\Theta} = & \sum_{n=1}^\infty \int \ldots \int K_{l,n}(\tau_1 \ldots \tau_n; \beta_{i,l,n}) 
\prod_i^n x(t_l^O - \tau_i) d\tau_i 
+ \sum_k \eta(t_l^O - t_k^O; \mu)
\end{split}
\end{equation}

\subsection*{Theoretical overview of STD}

Spike-triggered descent updates parameters based off of the distance between simulated and desired spike trains.  To support using the distance \eqref{eq:E} from \cite{banerjee2016learning} on the GCSRM, it is important to also generalize spike trains and show that they're a subset of a vector space with an inner product $\langle\cdot,\cdot\rangle$.  Considering augmented spike trains with countably infinite ($\mathbb{N}=\{1,2,...\}$) spikes gives the framework the versatility to compare spike trains of any length.  The augmentation turns spike trains into tuples of times and coefficients $\bm{t} = \{(t_i, \alpha_i)\}$.  This forms a vector space and provides the foundation for creating an inner product \eqref{eq:IP} that induces a metric (Methods).  Setting the $\alpha 's$  to $1$ for finitely many spikes reduces to the usual space of spike trains within a bounded past.  This is a subset of the generalization with the same metric and is squared to simplify the algebra \eqref{eq:distance}.

\begin{equation} \label{eq:IP}
\langle \bm{t^A}, \bm{t^B} \rangle = 
\sum_{i,j=1}^{\infty} (\alpha_i^A \times \alpha_j^B) 
\frac{t_i^A \times t_j^B}{(t_i^A + t_j^B)^2} 
e^{-\frac{t_i^A + t_j^B}{\tau}}
\end{equation}

\begin{equation} \label{eq:distance}
E = d^2 = \langle \bm{t^D} - \bm{t^O}, \bm{t^D} - \bm{t^O} \rangle
\end{equation}

Setting $\alpha$ values to 1 for $N,M$ spikes, and 0 otherwise, we get:
 
\begin{equation} \label{eq:E}
\begin{split}
E( \bm{t^D}, \bm{t^O} ) = 
&\sum_{i,j=1}^{M,M} \frac{t_i^D \times t_j^D}{(t_i^D+t_j^D)^2} e^{- \frac{t_i^D + t_j^D}{\tau}}
\\&+
\sum_{i,j=1}^{N,N} \frac{t_i^O \times t_j^O}{(t_i^O+t_j^O)^2} e^{- \frac{t_i^O + t_j^O}{\tau}}
\\&-2
\sum_{i,j=1}^{M,N} \frac{t_i^D \times t_j^O}{(t_i^D+t_j^O)^2} e^{- \frac{t_i^D + t_j^O}{\tau}}
\end{split}
\end{equation}

\begin{figure}[h!]
\begin{subfigure}{0.25\textwidth}
\scalebox{0.45}{
%This uses both pdf and pdf_tex
%This is so that the text size can be changed
\LARGE
%% Accompanies image file 'vector_space_scale3.pdf' (pdf, eps, ps)
\begingroup%
  \makeatletter%
  \providecommand\color[2][]{%
    \errmessage{(Inkscape) Color is used for the text in Inkscape, but the package 'color.sty' is not loaded}%
    \renewcommand\color[2][]{}%
  }%
  \providecommand\transparent[1]{%
    \errmessage{(Inkscape) Transparency is used (non-zero) for the text in Inkscape, but the package 'transparent.sty' is not loaded}%
    \renewcommand\transparent[1]{}%
  }%
  \providecommand\rotatebox[2]{#2}%
  \newcommand*\fsize{\dimexpr\f@size pt\relax}%
  \newcommand*\lineheight[1]{\fontsize{\fsize}{#1\fsize}\selectfont}%
  \ifx\svgwidth\undefined%
    \setlength{\unitlength}{283.46456693bp}%
    \ifx\svgscale\undefined%
      \relax%
    \else%
      \setlength{\unitlength}{\unitlength * \real{\svgscale}}%
    \fi%
  \else%
    \setlength{\unitlength}{\svgwidth}%
  \fi%
  \global\let\svgwidth\undefined%
  \global\let\svgscale\undefined%
  \makeatother%
  \begin{picture}(1,1)%
    \lineheight{1}%
    \setlength\tabcolsep{0pt}%
    \put(0.77026825,0.22666718){\color[rgb]{0,0,0}\makebox(0,0)[lt]{\lineheight{1.25}\smash{\begin{tabular}[t]{l}$\bm{t^A}+\bm{t^B}$\end{tabular}}}}%
    \put(0.88602348,0.7278624){\color[rgb]{0,0,0}\makebox(0,0)[lt]{\lineheight{1.25}\smash{\begin{tabular}[t]{l}$2 \bm{t^A}$\end{tabular}}}}%
    \put(0.3954158,0.6015773){\color[rgb]{0,0,0}\makebox(0,0)[lt]{\lineheight{1.25}\smash{\begin{tabular}[t]{l}$\bm{t^A}$\end{tabular}}}}%
    \put(0.3908683,0.3540332){\color[rgb]{0,0,0}\makebox(0,0)[lt]{\lineheight{1.25}\smash{\begin{tabular}[t]{l}$\bm{t^B}$\end{tabular}}}}%
    \put(0,0){\includegraphics[width=\unitlength,page=1]{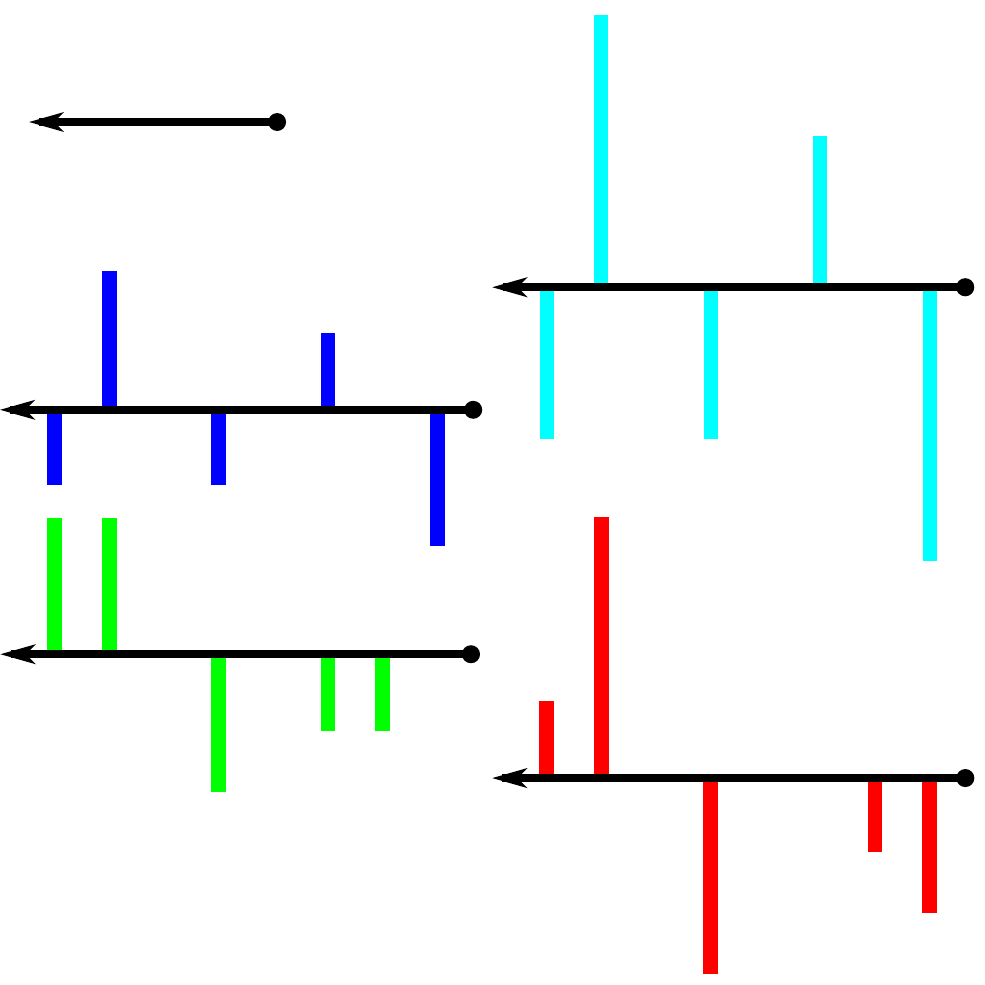}}%
    \put(0.06614583,0.88676758){\color[rgb]{0,0,0}\makebox(0,0)[lt]{\lineheight{1.25}\smash{\begin{tabular}[t]{l}Past\end{tabular}}}}%
  \end{picture}%
\endgroup%

}
\caption{\label{fig:spike_vec_linearity} }
\end{subfigure}
\begin{subfigure}{0.75\textwidth}
\input{./tex/results-gradient-matrix.tex}
\caption{ \label{fig:Dtdb} }
\end{subfigure}
\caption{ \textbf{a}, Two example spike trains $\bm{t^A}$ (blue) and $\bm{t^B}$ (green) under scalar multiplication $2\bm{t^A}$ (cyan) and vector addition $\bm{t^A} + \bm{t^B}$ (red). \textbf{b}, Example vectorization/memoization \cite{michie1968memo} for calculating the total derivatives of time changes with respect to changes in $\beta$ \eqref{eq:Dtdb} for spline $i$ at each prior spike.  }
\end{figure}

A diagram of the generalized version of spikes is shown in Figure \ref{fig:spike_vec_linearity}.  It illustrates scalar multiplication (cyan) and vector addition (red).  The vectors point positively into the past from the current time at zero.  Multiplying the vector $\bm{t^A}$ by $2$ has the effect of doubling the coefficient for each time (cyan).  When vectors $\bm{t^A}$ and $\bm{t^B}$ are added, equal times cause their coefficients to be added, else concatenated.  Coefficients which sum to zero cause the spike to be deleted.

%\subfile{./tex/spline.tex}

The GCSRM \eqref{eq:GCSRM} describes when the threshold is equal to the sum of the convolution(s) minus past AHP effects.  Perturbing its parameters, performing a first order Taylor approximation, and setting the perturbed and unperturbed expressions equal allows for the creation of partial derivatives with respect to the parameters $\beta$ \eqref{eq:dtdb}, $\mu$ \eqref{eq:dtdu}, and $t_k$ \eqref{eq:dtdt} (Methods).  These express how spike times change with respect to a change in spline coefficient, time constant, and prior spike time.  These partial derivatives are used to calculate the total derivatives \eqref{eq:Dtdb}, an example for $4$ spikes is shown in \ref{fig:Dtdb}.  The matrix for $\mu$ is similar except that $\mu$ only effects future spikes meaning $\frac{\partial t_l}{\partial \mu_l} = 0$.  The total derivative for a spike time with respect to a parameter at a prior time is the sum of that parameter's total effect on each of the more recent prior spikes each multiplied by the relevant time partial to tie it to the current time.

\begin{equation} \label{eq:dtdb}
\frac{\partial t_l^O}{\partial \beta_{i,l}} 
= 
\frac{
- \int B_i(\tau) x(t_l^O - \tau) d\tau 
}{
\int_0^{|K|} K(\tau; \beta_{i,l}) \frac{\partial x}{\partial t} \Big|_{t_l^O - \tau} d\tau 
+ \sum \frac{\partial \eta}{\partial t} \Big|_{t_l^O - t_k^O}
}
\end{equation}

\begin{equation} \label{eq:dtdu}
\frac{\partial t_l^O}{\partial \mu_l} 
= 
\frac{
- \frac{\partial \eta}{\partial \mu} \Big|_{t_l^O - t_k^O}
}{
\int_0^{|K|} K(\tau; \beta_{i,l}) \frac{\partial x}{\partial t} \Big|_{t_l^O - \tau} d\tau 
+ \sum \frac{\partial \eta}{\partial t} \Big|_{t_l^O - t_k^O}
}
\end{equation}

\begin{equation} \label{eq:dtdt}
\frac{\partial t_l^O}{\partial t_k^O} 
= 
\frac{
\frac{\partial \eta}{\partial t} \Big|_{t_l^O - t_k^O} }{
\int_0^{|K|} K(\tau; \beta_{i,l}) \frac{\partial x}{\partial t} \Big|_{t_l^O - \tau} d\tau 
+ \sum \frac{\partial \eta}{\partial t} \Big|_{t_l^O - t_k^O}
}
\end{equation}
\begin{equation} \label{eq:Dtdb}
\frac{D t_{k+1}^O}{\partial \beta_{i,l}} 
= 
\sum_{t_j > t_l}^{k}
\frac{D t_j^O}{\partial \beta_{i,l}}
\frac{\partial t_{k+1}^O}{\partial t_j^O}
\end{equation}

The total derivative expressions for each of the spikes with respect to a parameter are then connected via another chain rule to the squared distance E.  This is performed by summing over the multiple of $\frac{\partial E}{\partial t_{k}^O}$ along a (colored) diagonal in \ref{fig:Dtdb}.  The sum of all of these changes in E with respect to that parameter across all time steps is the gradient.  The equations \eqref{eq:db} and \eqref{eq:du} are the gradient updates used to minimize the spike train distance with respect to the spline parameters $\beta$ and AHP time constant $\mu$.  Using these gradients, small steps (with size $\alpha$) are taken in accordance with sgd methods such as momentum \cite{qian1999momentum} and katyusha \cite{allen2017katyusha}.  The updates are best triggered soon after spikes since E exponentially decays as spikes are pushed further into the past.

\begin{equation} \label{eq:db}
\beta_i = \beta_i - \alpha_\beta \sum_{k \in F_l}
\frac{\partial E}{\partial \beta_{i,l}}
\end{equation}

\begin{equation} \label{eq:du}
\mu = \mu - \alpha_\mu \sum_{k \in F_l}
\frac{\partial E}{\partial \mu_k}
\end{equation}

\biblio

\subsection*{Convergence of $\beta$ and $\mu$}

\begin{figure}
\begin{subfigure}{0.5\textwidth}
%  06_paired_1D_L2
\includegraphics[scale=0.5]{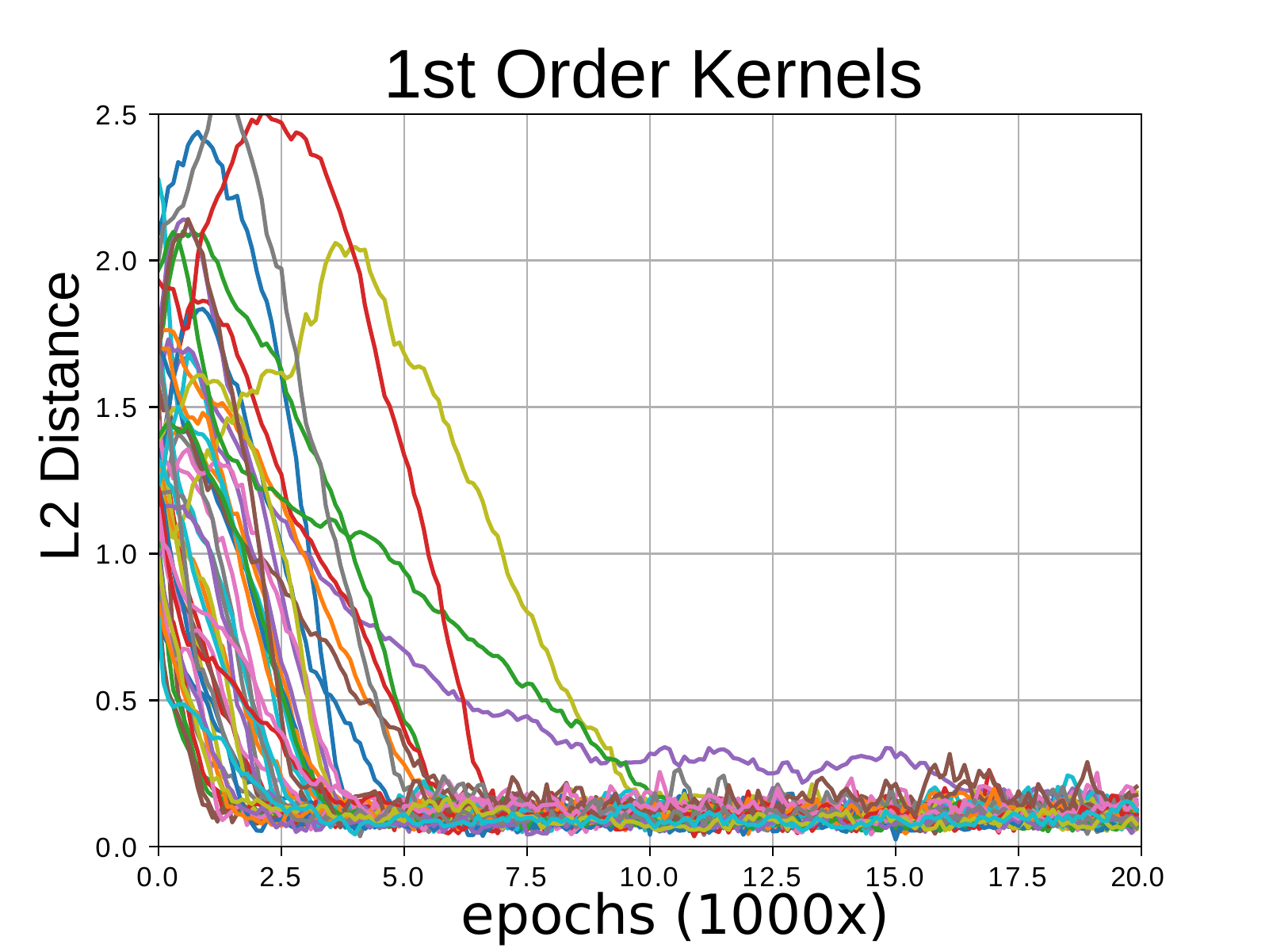}
 \caption{\label{fig:L2_50_1D}}
\end{subfigure}%
\begin{subfigure}{0.5\textwidth}
%  09_2D_L2
\includegraphics[scale=0.5]{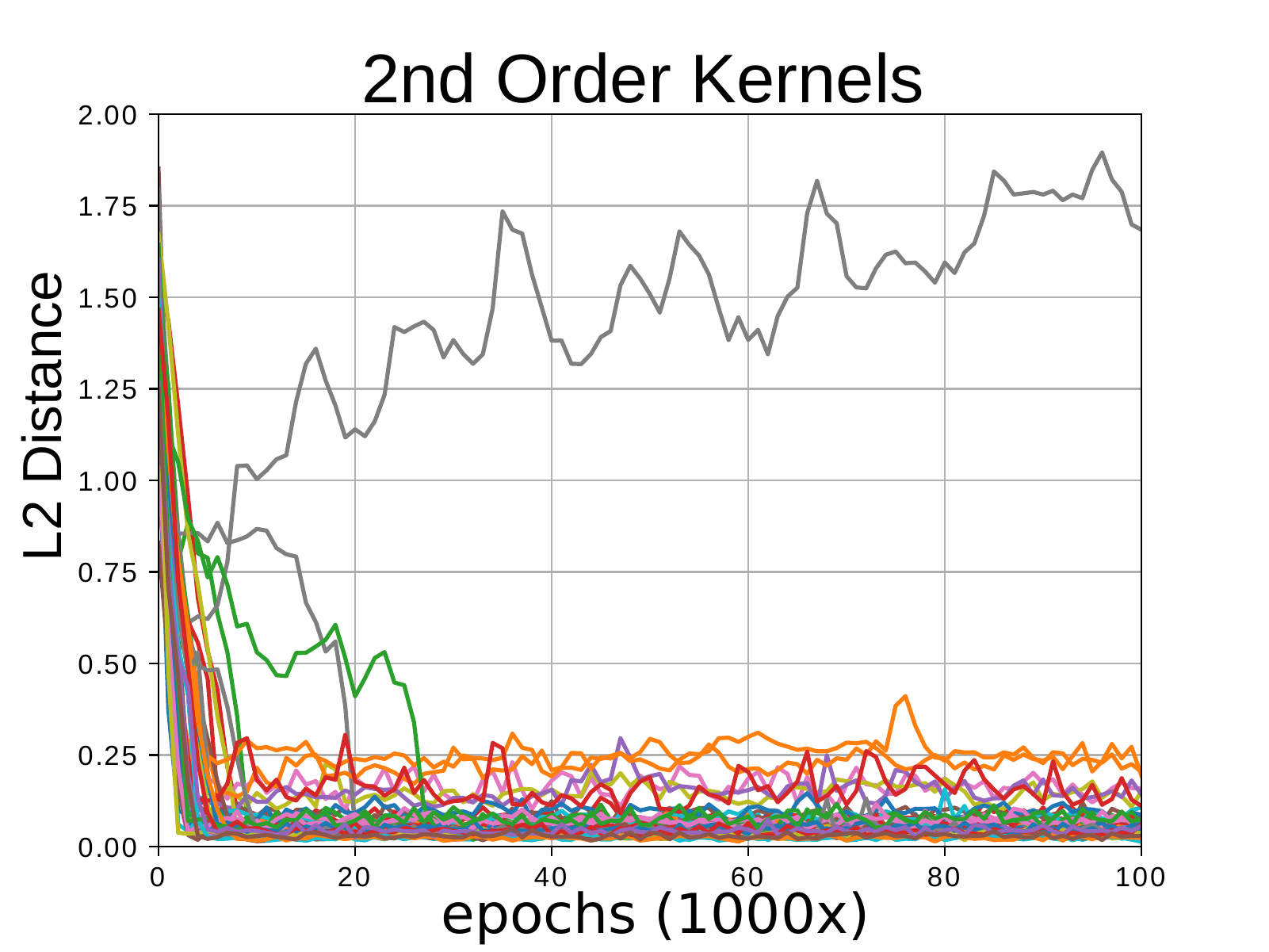}
 \caption{\label{fig:L2_50_2D}}
\end{subfigure}%
 \caption{\label{fig:convergence} \textbf{a},\textbf{b}, Convergence for $50$ trials of 1st order kernels composed of $10$ splines and 2nd order kernels composed of a grid of $8 \times 8$ splines.  Each epoch represents an update and consists of a signal of $400$ time units set as $1ms$ meant to be consistent with a statically set AHP time constant $\mu = 1.2ms$. }
\end{figure}

Convergence of first and second order kernels on simulated data are shown in Figure \ref{fig:convergence}.  While holding $\mu$ fixed, the convergence for 50 trials of 1st (\ref{fig:L2_50_1D}) and 2nd (\ref{fig:L2_50_2D}) order kernels are demonstrated.  An error, or $L_2$ distance, of zero means the learning and desired kernels are equal.  A variety of SGD methods were used: vanilla was too slow, katyusha was too unstable, and momentum was fast and relatively robust.

One simulation, seen in (\ref{fig:L2_50_2D}), does not converge.  These cases can arise for a variety of reasons and is often the result of hyperparameters: momentum, learning rate, cap, AHP threshold, and number of epochs.  Lower learning rates and momentum converge more stably but require more iterations.  For 2nd order kernels, a mask is applied to keep it lower triangular.  This was used as a means of preventing diagonally symmetric terms from conflicting.  The cap, a crucial parameter due to the highly nonlinear nature of the problem, is an upper bound on the $L_2$ norm of the gradient step preventing unnecessarily large updates which lead to divergence.

When the desired kernel is shorter than the learning kernel, the extra values become zero as would be expected.  Convergence was best for these third order overlapping cardinal B-splines with $12$ abstract time units of nonzero support.  Most tests were performed on $10$ splines.  More splines took more training time and this was compounded by increasing the kernel's order.  The computational and space requirements of higher order kernels restrict consideration to fewer parameters.

\begin{figure}
\begin{subfigure}{0.5\textwidth}
\includegraphics[scale=0.385]{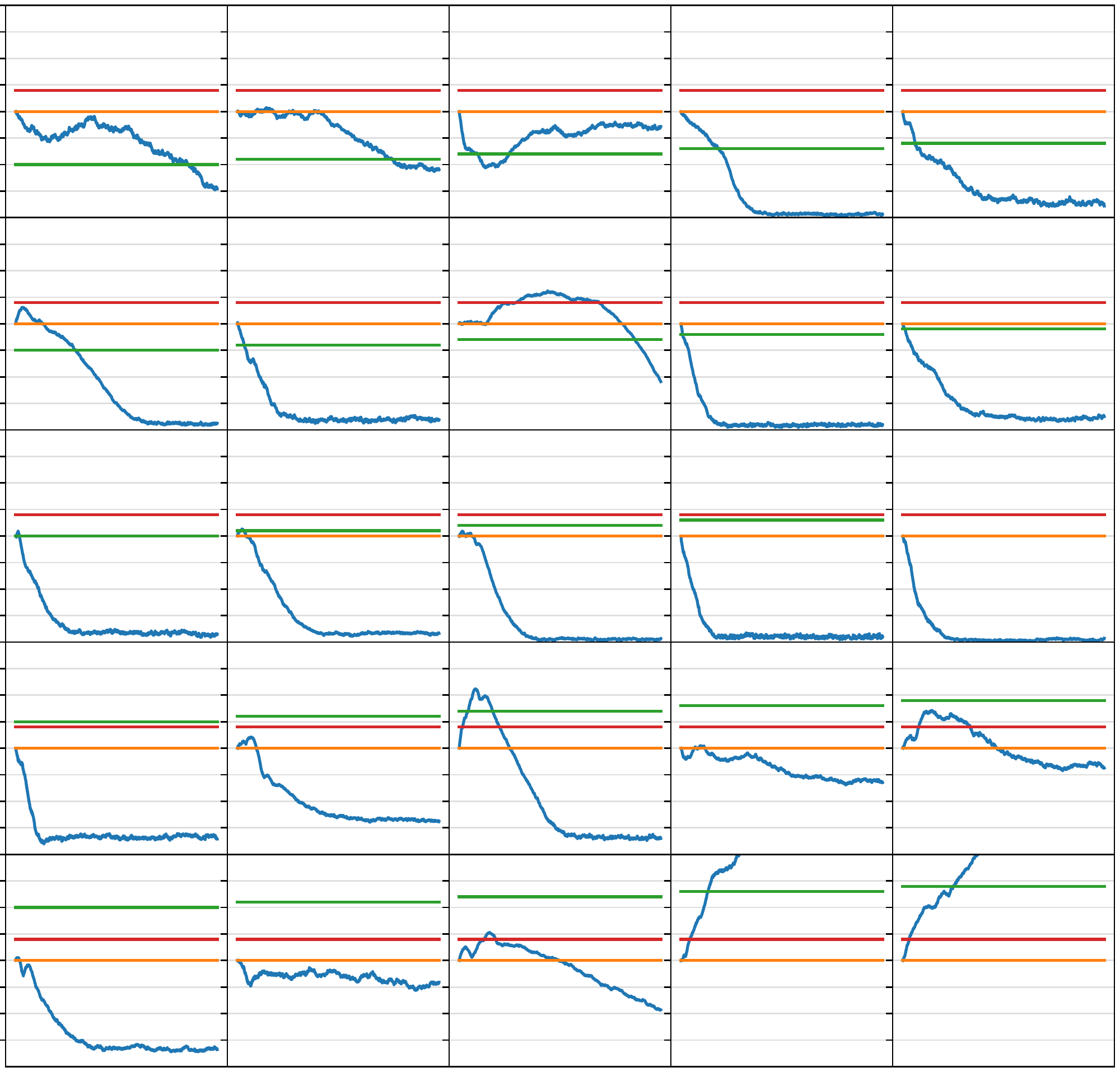}
 \caption{\label{fig:mu_wrong_fixed}}
\end{subfigure}
\begin{subfigure}{0.5\textwidth}
\includegraphics[scale=0.385]{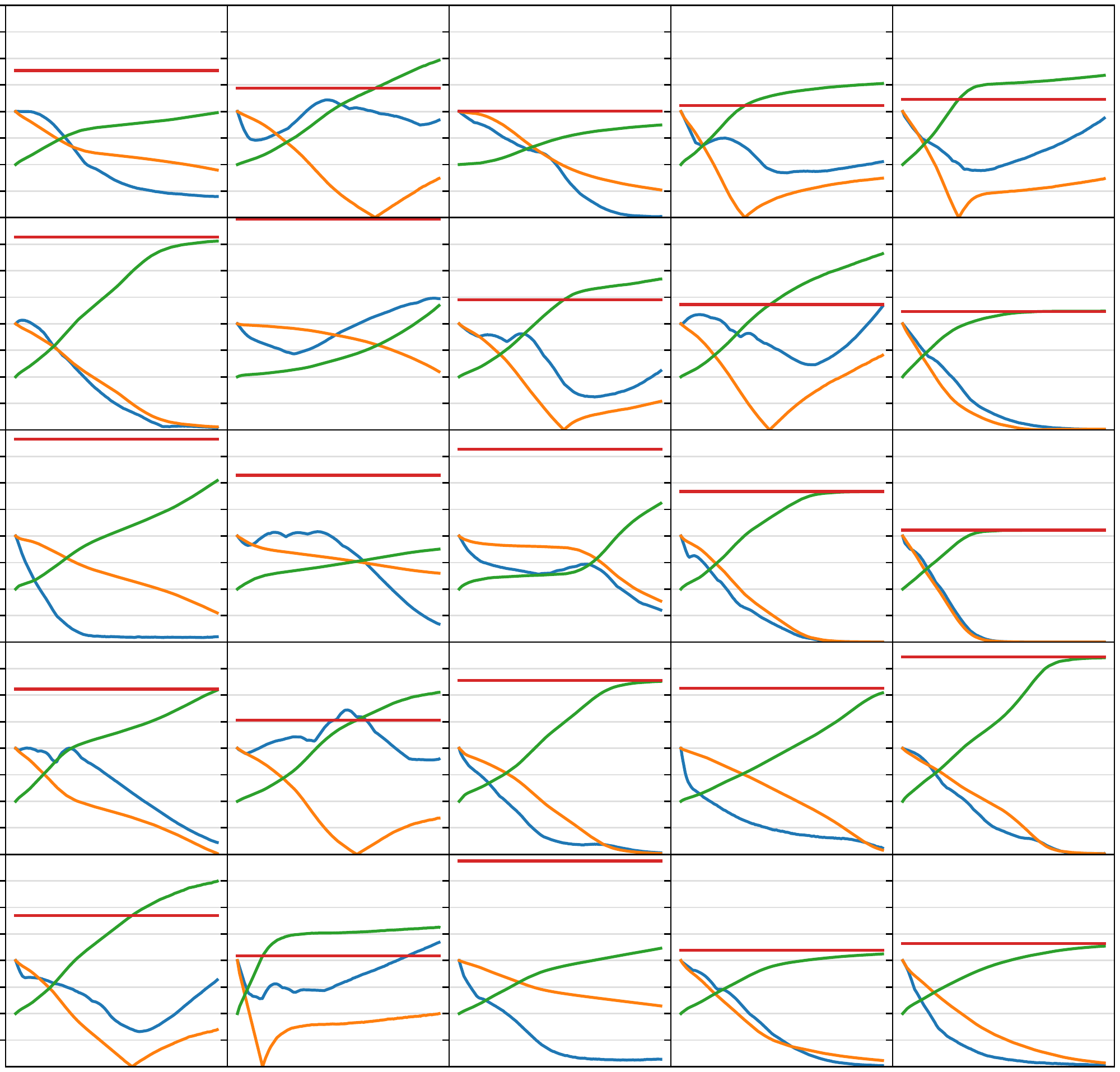}
 \caption{\label{fig:both_learning}}
\end{subfigure}
 \caption{ \label{fig:mu} \textbf{a}, $\mu$ (green) is scanned from $0.50$ (top left) to $1.7$ (bottom right) in steps of $0.05$.  For a desired $\mu = 1.2$ (red) a range for accurate convergence for $\beta$ (blue) corresponds to $[0.65, 1.35]$ or alternatively $[54\%, 113\%]$ of the desired. \textbf{b}, Convergences are shown for $\mu$ (orange) and $\beta$ (blue) being simultaneously learned and plotted as ratios from their original $L_2$ distance. }
\end{figure}

In Figure \ref{fig:mu_wrong_fixed}, the learning $\mu$ (green) is set incorrectly from the desired $\mu$ (red) and scanned for its effect on the convergence of the $\beta$ parameters.  There is a larger region of stability below the desired $\mu$.  The $\mu$ (orange) and $\beta$ (blue) errors are normalized with respect to their initial distance from the desired.  Demonstrated in \ref{fig:both_learning}, $\mu$ and $\beta$ are being learned in tandem.  Once the learning $\mu$ grows larger than the desired, the simulations begin to diverge.  This is because $\mu$'s effect is exponential and $\beta$'s splines are linear.  They cannot compensate for this error and this causes a feedback loop.  Seeing as there was a large range of stability it is something that can easily be scanned.  We also tried using a spline representation for the AHP, but getting the parameters to match required spikes to overlap with the spline.  Some splines, particularly immediate after a firing, were the least likely to occur but also had the greatest impact.  Other basis functions might be an option.

\biblio

\subsection*{Sensitivity to Noise: shifting addition and deletion of spikes}

\begin{figure}
\begin{subfigure}{.5\textwidth}
\includegraphics[scale=0.5]{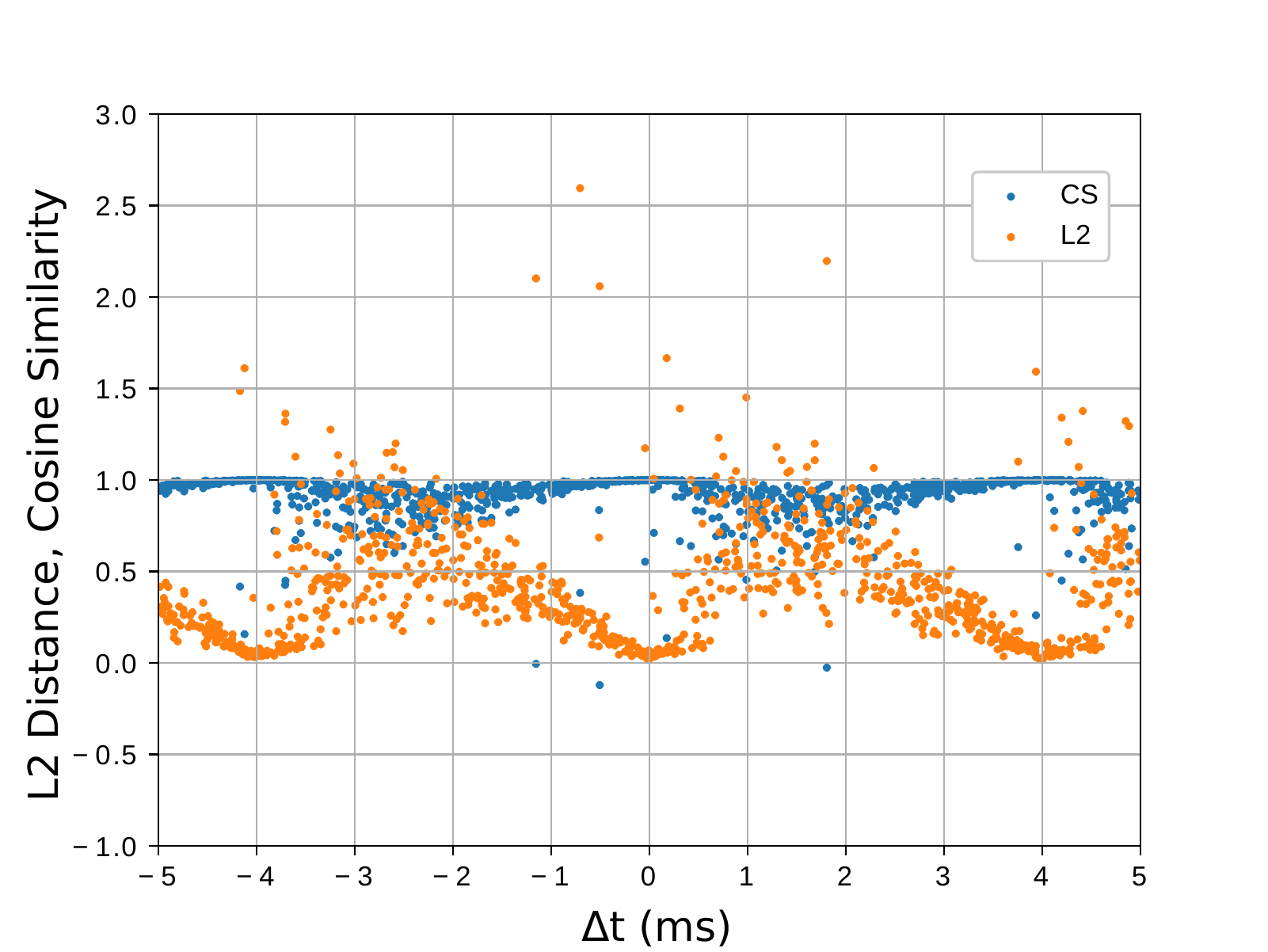}
 \caption{ \label{fig:Noise_dt} }
\end{subfigure}%
\begin{subfigure}{.5\textwidth}
\includegraphics[scale=0.5]{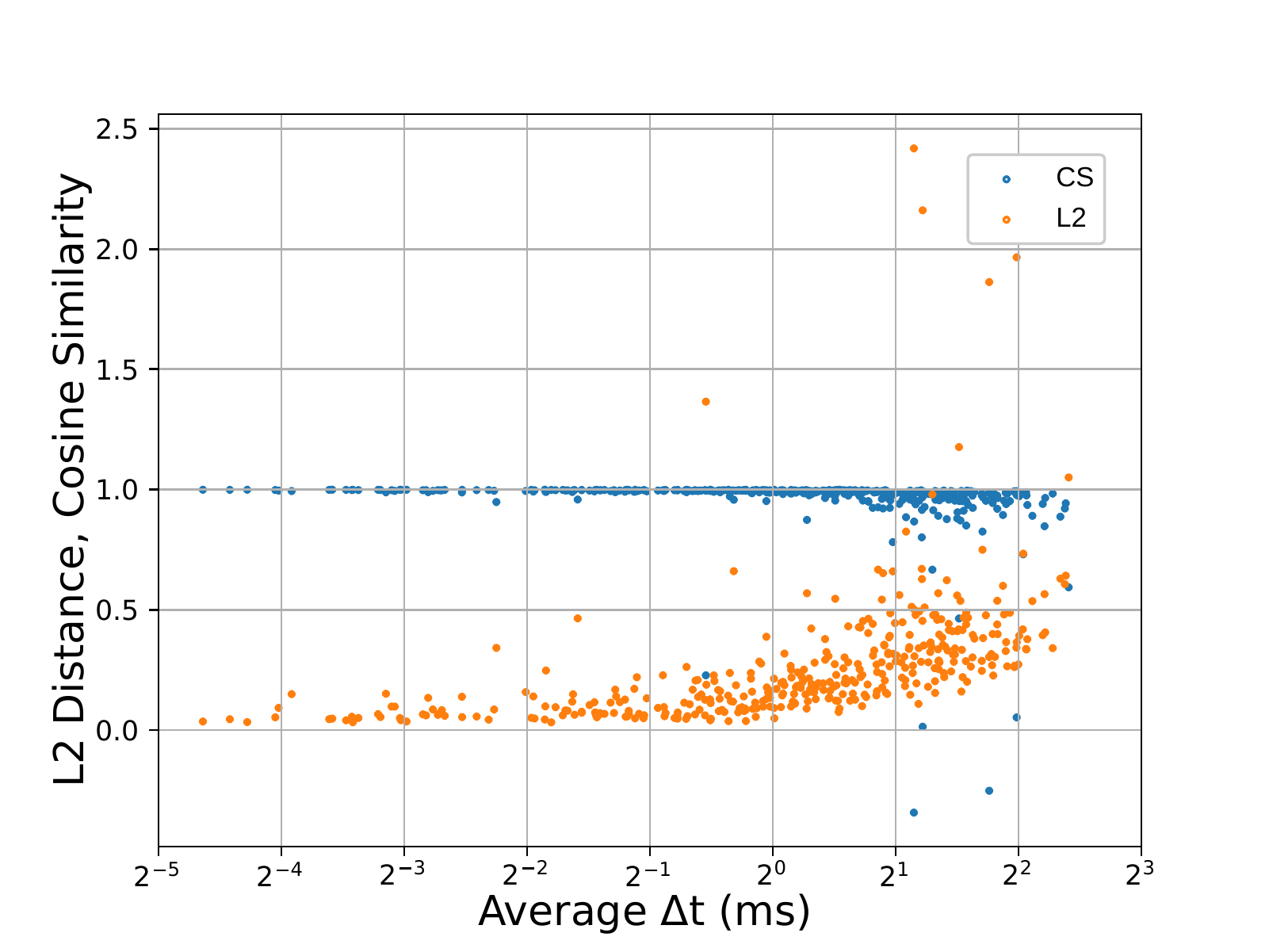}
 \caption{ \label{fig:Noise_dtn} }
\end{subfigure}%

\medskip

\begin{subfigure}{.5\textwidth}
\includegraphics[scale=0.5]{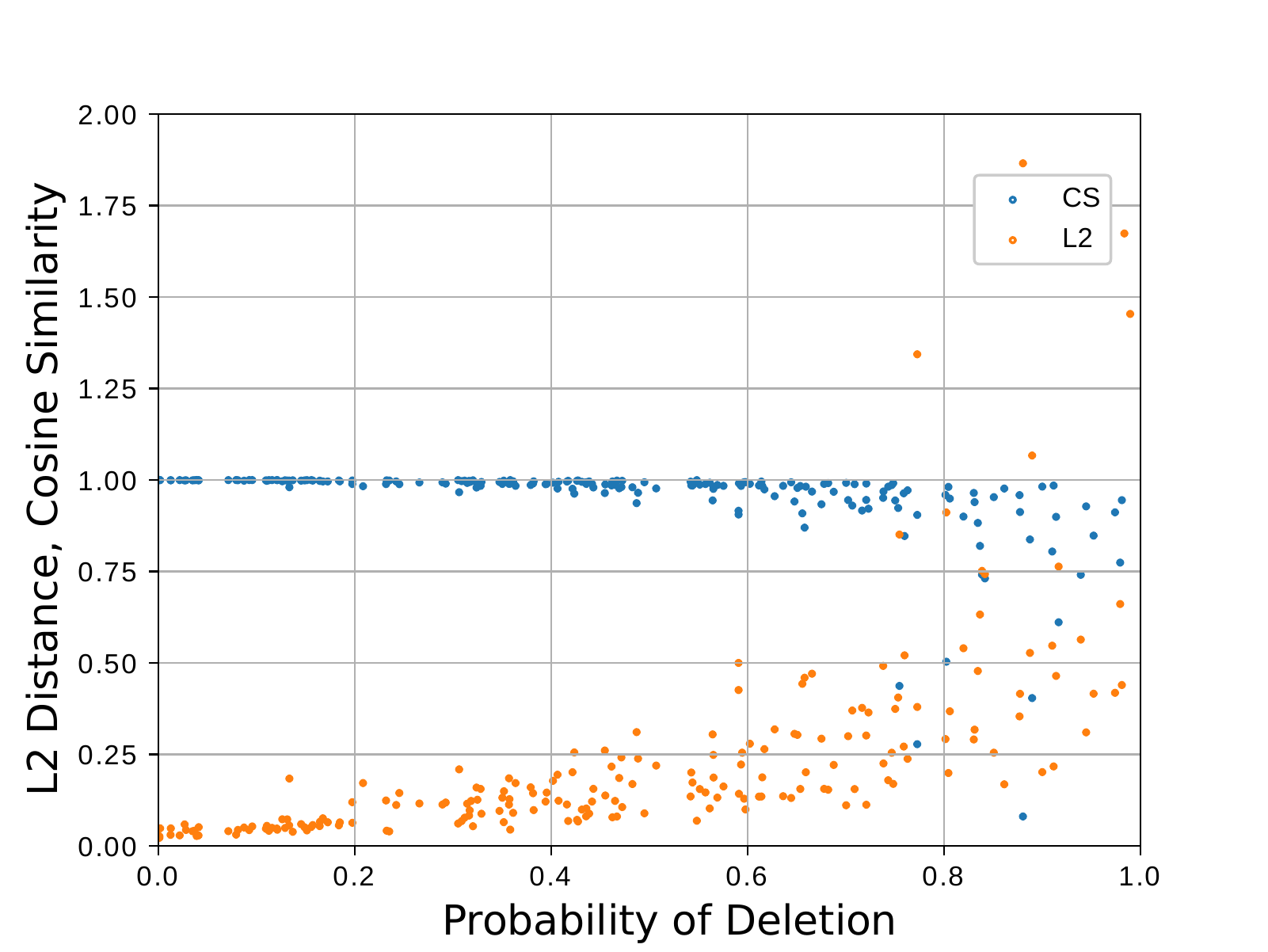}
 \caption{ \label{fig:Noise_Del} }
\end{subfigure}%
\begin{subfigure}{.5\textwidth}
\includegraphics[scale=0.5]{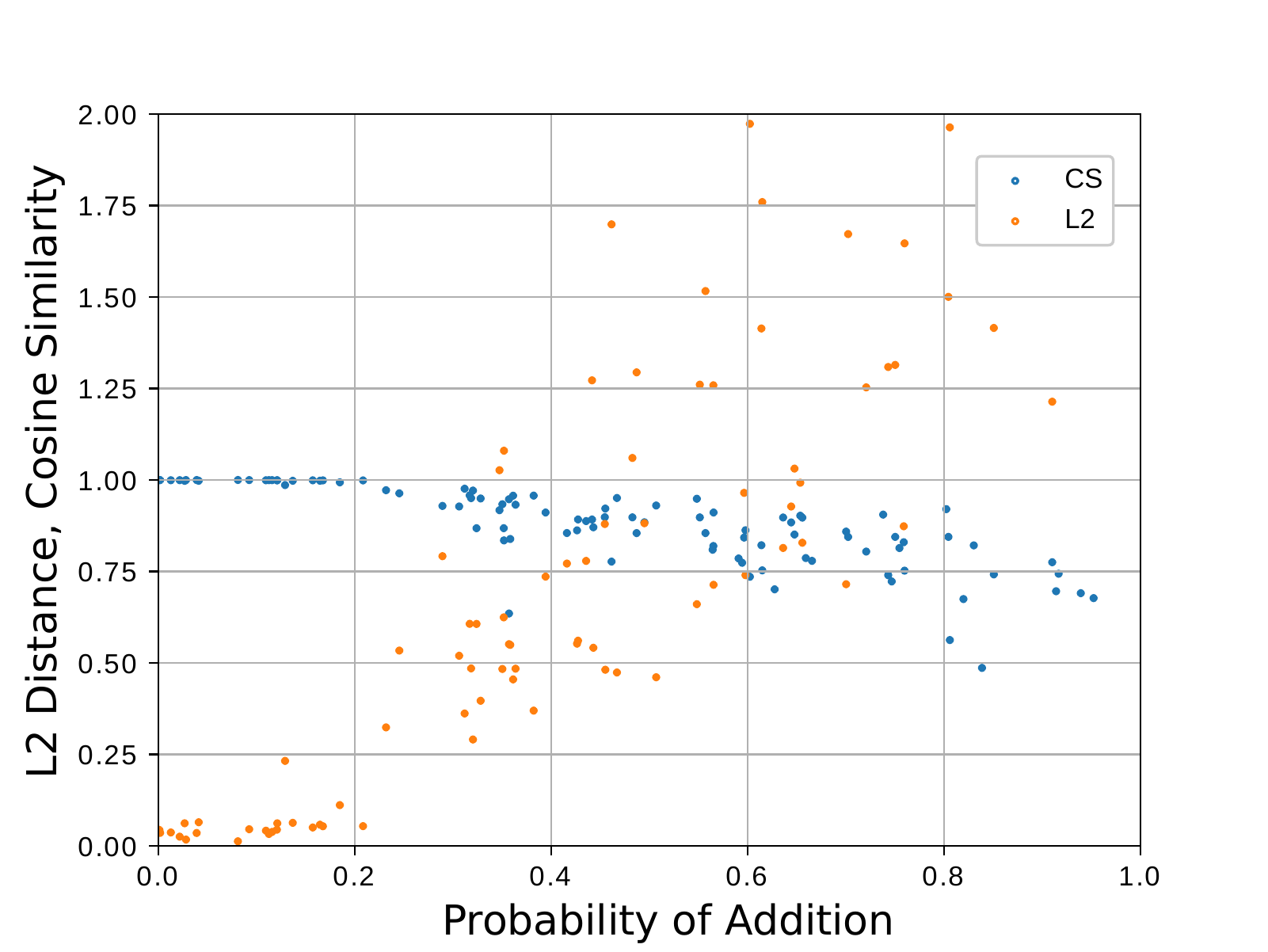}
 \caption{ \label{fig:Noise_Add} }
\end{subfigure}%
 \caption{ \label{fig:Noise} \textbf{a}, Shifting spikes uniformly by a set amount resulted in a cyclic pattern as the $10$ splines snapped into alignment with adjacent splines. \textbf{b},  Randomly perturbing each desired spike separately about a normal distribution. \textbf{c}, \textbf{d}, Picking a homogeneous Possion rate for deleting (or adding) spikes reveals that as many as half (or a fifth) of the spikes could be deleted (or added) while still extracting a kernel with a low $L_2$ distance.  }
\end{figure}

After giving STA difficulty with a complex AHP scenario, it was important to similarly challenge the capability of STD by from learning LNP spikes.  This exploration was initiated by analyzing the effect on convergence by the shifting, adding, or deleting of spikes.  Upon perturbing the spike times by some fixed amount ($dt$) it quickly became apparent that more granularity was needed in the AHP spike generation process itself.  This lead to calculating a more precise intersection of the potential with the threshold and was achieved by linear interpolation.  Figure \ref{fig:Noise} demonstrates the effects of adding noise to the CSRM system and the effect it has on convergence and cosine similarity defined by $\frac{a \cdot b}{||a|| \; ||b||}$ which is chosen for its invariance to scaling.

Moving far away from $dt=0$ resulted in divergence and indicated that the learning kernel needed flexibility.  Padding both the learning and desired kernels with splines having $\beta = 0$ allowed the learning kernel to appropriately shift (\ref{fig:Noise_dt}).  Removing spikes (\ref{fig:Noise_Del}) randomly with some probability, normalized with respect to the number of spikes, had minimal effect on convergence until the deletion chance was $~50\%$.  Real neurons are highly connected and auxiliary processing could cause additional seemingly random spikes.  Adding spikes to any location (\ref{fig:Noise_Add}) shows that learning is comparatively more sensitive to addition. 

%%%%%%%%%%%%%%%
%Noise Diagrams
%%%%%%%%%%%%%%%

%%%%%%%%%%%%%%%
%All and 3rd Worst LNP
%%%%%%%%%%%%%%%
\begin{figure}[!htb]
    \centering
    \begin{tabular}[t]{cc}
\begin{subfigure}{0.5\textwidth}
    \centering
    \smallskip
    \includegraphics[width=1.1\linewidth,height=1.0\textwidth]{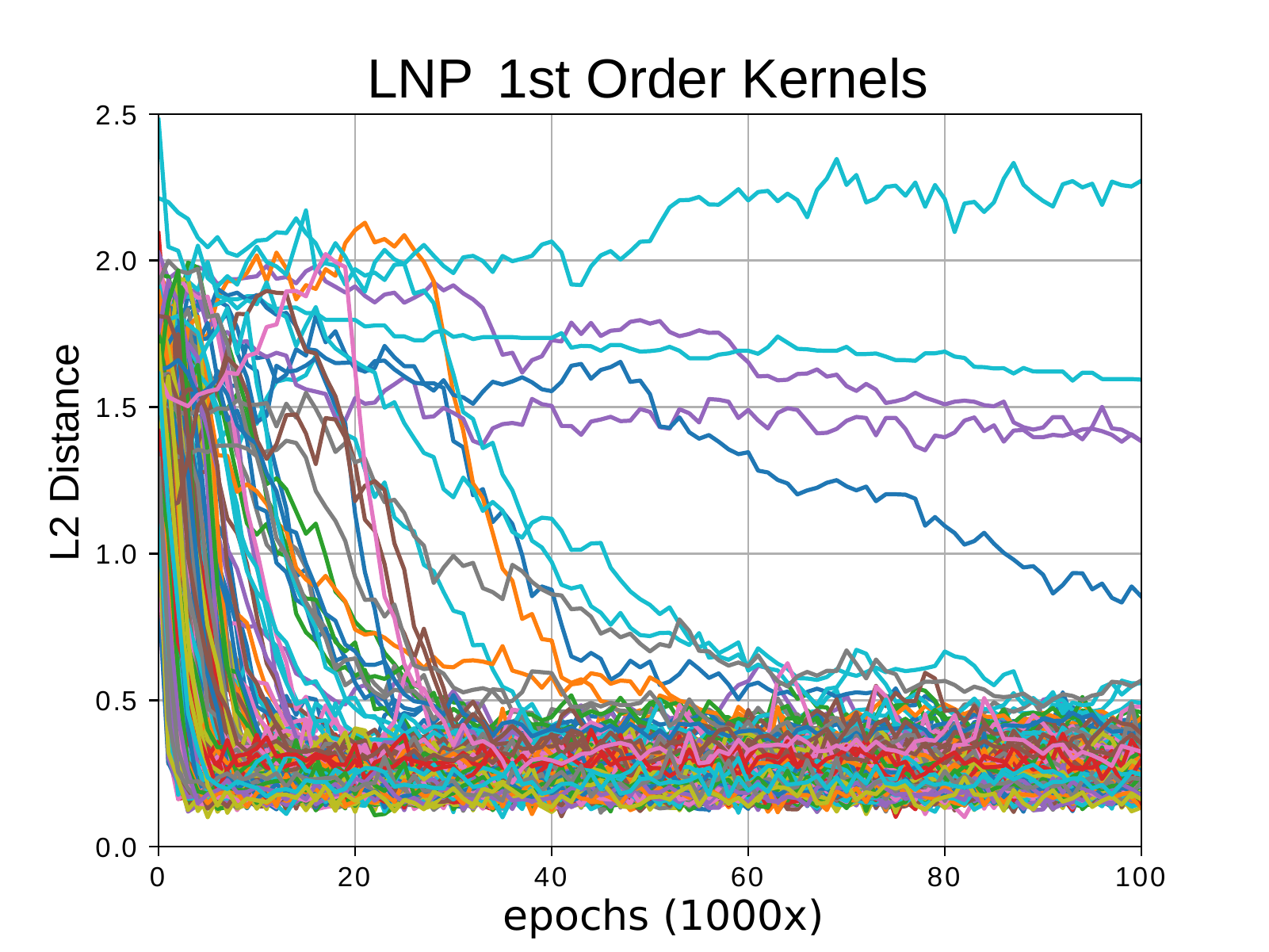}
    \caption{}
\end{subfigure}
    &
        \begin{tabular}{c}% if you add [t], than sub images are pushed down
        \smallskip
            \begin{subfigure}[t]{0.4\textwidth}
                \centering
                \includegraphics[width=0.9\textwidth,height=0.4\textwidth]{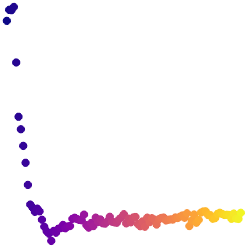}
                \caption{}
            \end{subfigure}\\
            \begin{subfigure}[t]{0.4\textwidth}
                \centering
                \includegraphics[width=0.9\textwidth,height=0.4\textwidth]{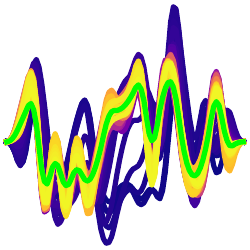}
                \caption{}
            \end{subfigure}
        \end{tabular}\\
    \end{tabular}
\caption{\label{fig:LNP}  \textbf{a}, Convergence for $500$ STD trials on a $4$-binned LNP process learning $20$ splines of $80ms$ over $100,000$ epochs of length $400ms$.  \textbf{b}, Thirteenth worst convergence peaking at an $L_2$ distance of $178\%$ and ending at $42.3\%$. \textbf{c}, Intermediary learning kernels vs the desired (lime).}
\end{figure}

STD's success on spikes generated via an LNP process is explored in Figure \ref{fig:LNP}.  For the LNP model, a high exponential coefficient in the nonlinearity (a sigmoid) resulted in a firing rate which was mostly either $0$ or $1$.  This made it crucial to center it around the threshold value.  This resulted in too many spikes above threshold and none below.  To obtain a corresponding amount of spikes in both the LNP and AHP models a mixture of modifications can be made: reducing $\mu$ the AHP time constant, reducing the peak probability of spiking, and constraining the LNP time domain.  The peak probability was set at $50\%$ and the LNP spike time domain was binned down sampling by a factor of $4$.  Among the $500$ cases, the average learning kernel had a $20\%$ lower error.

\biblio

\subsection*{Kernel for \textit{Locusta migratoria} tympanal nerve}
Sensory neurons which, when given the same stimuli, reliably produce similar spike trains are good candidates to test STD.  One such possibility is the tympanal nerve's auditory receptor axons in the \textit{Locusta migratoria} grasshopper whose action potentials, recorded intracellularly, can have a $0.15ms$\cite{rokem2006spike} inter trial jitter.  This dataset, collected by Ariel Rokem\cite{eyherabide2009bursts} \cite{eyherabide2008burst} at the lab of Andreas Herz, was graciously shared through the CRCNS program (http://crcns.org) \cite{https://doi.org/10.6080/k0bg2kwb}.   The stimuli consisted of a carrier wave perturbed by random amplitude modulations and a cutoff frequency of up to $800Hz$.  We used a particular subset of this dataset to demonstrate kernel extraction.

%https://tex.stackexchange.com/questions/25414/how-to-create-magnified-subfigures-and-corresponding-boxes-for-portions-of-a-lar
\begin{figure}[ht]\centering
\begin{tikzpicture}[
    zoomboxarray,
    zoomboxarray columns=3,
    zoomboxarray rows=1,
%    connect zoomboxes,
    zoombox paths/.append style={line width=3pt}
]
    \node [image node] { \includegraphics[width=0.45\textwidth]{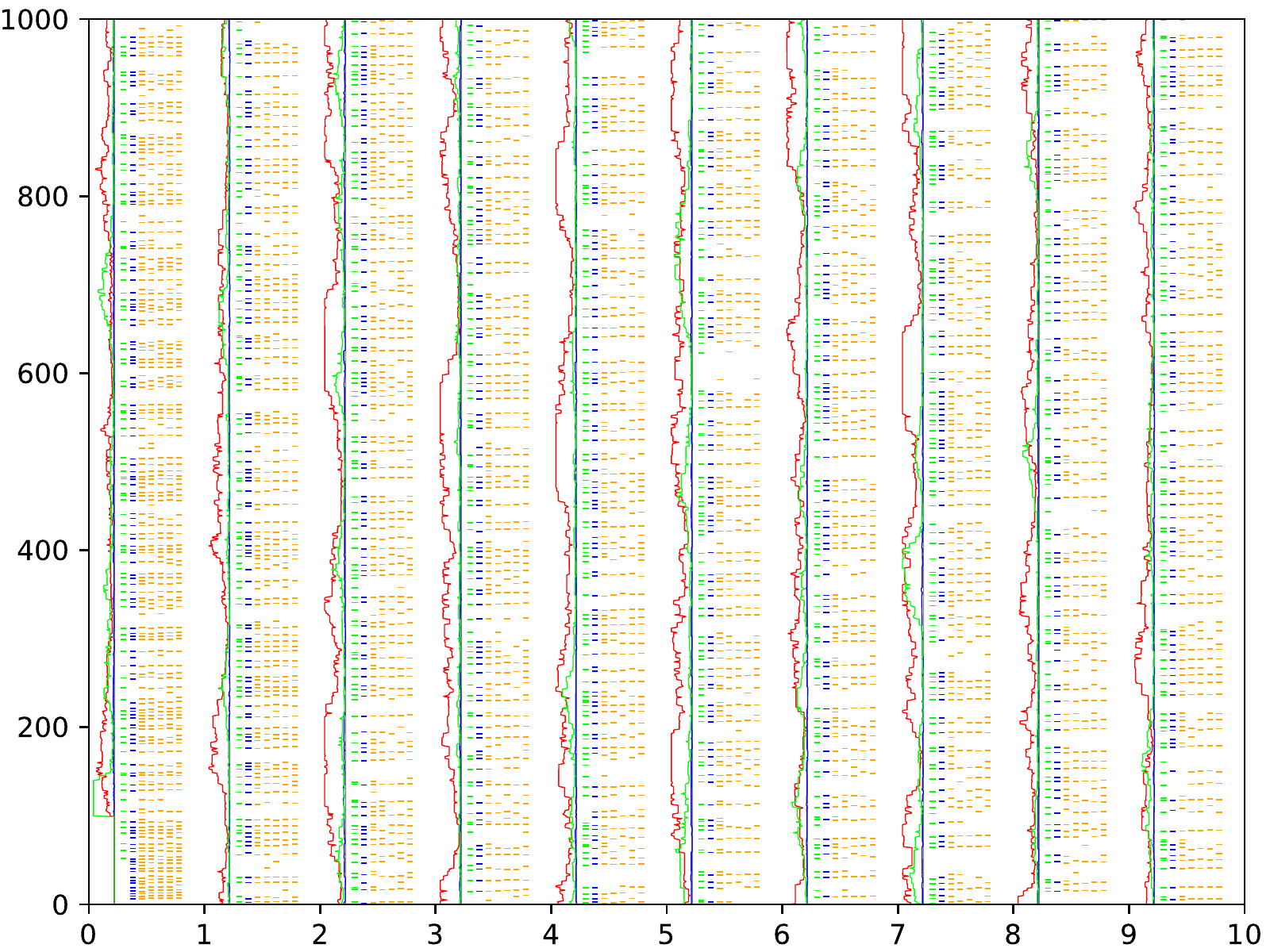} };
    \zoombox[magnification=4, color code=lime]{2*0.0911 + 0.108 , 7*0.075 + 0.18}
    \zoombox[magnification=4, color code=blue]{4*0.0911 + 0.108 , 4*0.075 + 0.18}
    \zoombox[magnification=4, color code=orange]{8*0.0911 + 0.108 , 3*0.075 + 0.18}
%    \zoombox[magnification=6, color code=purple]{3*0.0911 + 0.115 , 6*0.084 + 0.135}
%    \zoombox[magnification=6, color code=yellow]{8*0.0911 + 0.115 , 1*0.084 + 0.135}
\end{tikzpicture}
\caption{ \label{fig:50s5} \textbf{a}, The \textit{Locusta migratoria} spike data (blue and orange dots) collected by A. Rokem in cell ``./crcns-ia1/Data1/03-04-23-ad/" corresponding to the ``gauss\_st6\_co200.dat" stimuli with a cutoff frequency of $200Hz$.  The recording is sliced up with milliseconds on the y axis and seconds on the x for a total of $10$ seconds.  Each spike train consists of approximately $1000$ spikes.  The goal was to train an STD kernel that could reconstruct (green dashes) the desired (blue dashes) spike train.  There are $5$ other recordings (orange dashes) for this stimuli and cell.  The final kernel's simulated spike train closely approximates the desired.  The green line represents error between the learning and desired spike trains from a preceding window of $100ms$ and is clipped to a range from $0$ to $6$.  Similarly, the blue and red lines represent the minimum and maximum error between any pair of recorded spike trains.  \textbf{b}, Zoomed sections of the reconstruction were randomly chosen.  Notice that the learning error is often between the minimum and maximum errors between recordings.  Additionally, a simulated spike will sometimes align with the other recordings even when the desired does not.  These are indications that the kernel accurately represents underlying model. }
\end{figure}

Repeated updates lead to a kernel which attempts to reconstruct the provided spike train.  When running a simulation with a known desired neuron the absolute distance between kernels can be measured.  In absence of a known answer, and in lieu of a universally accepted metric, we revert to demonstrating the effectiveness by showcasing the simulated spike train reconstruction for a particular kernel.  Figure \ref{fig:50s5} demonstrates this using a (stimuli, spike train) pair as input.  The stimuli we used had a cutoff frequency of $200Hz$.  The reconstruction (green dashes) was learned from (blue dashes) one of the $6$ recorded spike trains for a particular cell (``./crcns-ia1/Data1/03-04-23-ad/").  The reconstruction occasionally fires where the desired does not but the other (orange dashes) recordings do.  This is an indication that the underlying model is being accurately represented.

\begin{figure}
\includegraphics[width=16cm, height=5cm, trim={2.4cm 1.6cm 2.65cm 1.8cm},clip]
{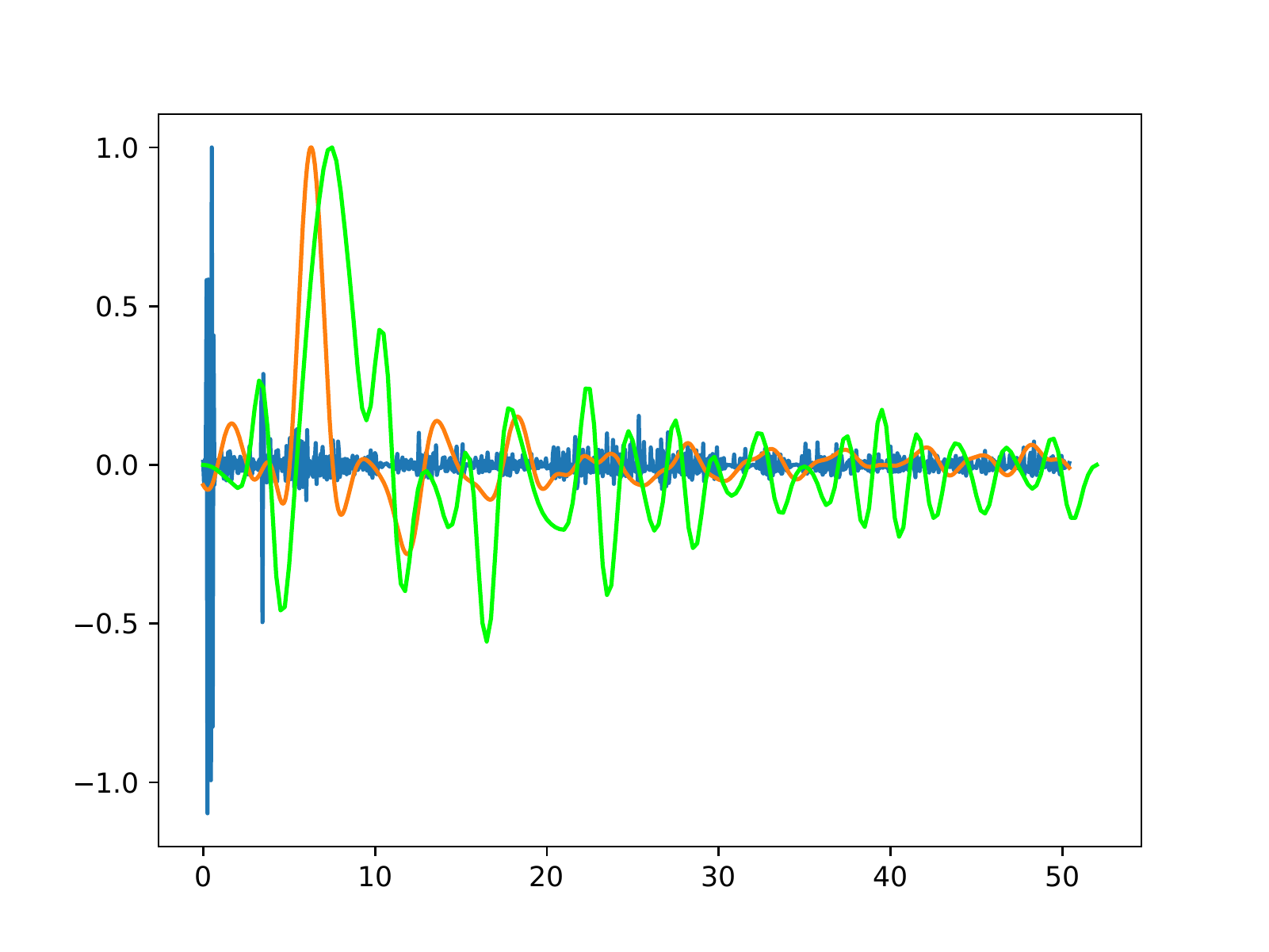}
\caption{\label{fig:LmKernels}  The raw STA kernel (blue) spanning $1000$ time units of $0.05ms$ was smoothed (orange) and compared with the STD result (green).  The smoothed STA kernel was the result of of convolving $5$ times against a box spline spanning $1ms$.}
\end{figure}

Given that the data set was recorded at $0.05ms$ and borrowing inspiration from Nemenman's work\cite{nemenman2008neural}, showing that a blowfly's H1 neuron represents information with sub-millisecond precision even for slow stimuli, finer time resolution STD kernels were explored.  The best observed reconstruction was for an STD kernel (green) composed of $50$ splines and four $0.25ms$ units per knot interval shown in Figure \ref{fig:LmKernels} along with raw (blue) and smoothed (orange) STA kernels.  The STA results were omitted from Figure \ref{fig:50s5} due to a large variety of low quality reconstructions.  It was trained with all $6$ recorded spike trains, different sized kernels, various levels of smoothing, and multiple spike generation methods (LNP, CSRM).  STD is a better technique for learning kernels that have the ability to reconstruct spike trains.

\biblio

\section*{Discussion}

STD, unlike STA, can learn higher order kernels and generate spike trains to verify the accuracy of its approximation.  STA is restricted to linear kernels and doesn't provide a nonlinearity for reconstruction.  STD provides a better kernel approximation than STA in our simulated examples.  Robustness to a variety of noise and a cross model comparison indicate its potential to be widely applicable.  The simulated reconstructions of the \textit{Locusta migratoria} spike trains are closely aligned which indicates that STD can extract the kernel from many different neural systems.

The primary concern is that model mismatch might make convergence impossible.  Unfortunately, this could be confused with incorrect hyperparameters such as the cap, AHP value, threshold, learning rate, momentum, and epochs.  Convergence can be measured in terms of parameter stabilization, the predictive ability of the simulation, and with the spike train distance.  Multiple simulations can be run from different initializations, but convergence to the same parameter set does not imply a correct solution.  This was first seen when trying to learn LNP spike trains where settings stably converged to an incorrect parameter set.  This was due to an incompatible model mismatch and was fixed by down sampling.  

Slowly changing kernels could be learned from slices of the signal and interpolating the intermediary results.  The implementation of the second order kernel operations are similar enough to first order such that core pieces of code can be left untouched for further generalization.   The lowest hanging fruit for this technique would be to apply it to other existing datasets.  Expanding the code to work for video inputs is another avenue for future research.

\biblio

\subsection*{Acknowledgments}
Anik Chattopadhyay and Daniel Crews for their assistance with theory and troubleshooting, Kyle Altendorf (@altendky) and Aleksi Torhamo (@Alexer) via \#python on IRC in freenode for help with profiling and vectorizing calculations.  And thanks to Ariel Rokem for sharing the \textit{Locusta migratoria} dataset and CRCNS for hosting it.  This work was partially funded by AFOSR grant \#FA9550-16-1-0135.

\biblio

\section*{Materials and Methods}

\subsection*{Spike Triggered Average as an optimization}

Ignoring the nonlinearity due to the proportionality via Bussgang's equation 20 \cite{bussgang1952crosscorrelation} leads to an optimization problem.  In the discrete time framework, each successive $|k|$ sized slice of a signal $x$ forms a row of a matrix $X$.  The convolution of the kernel with a corresponding section of the signal $k * x_w$ becomes $Xk$.  The resulting simulated electrical potential from the convolution is labeled $y$.  The optimization problem can then be phrased $min( || X k - y ||^2)$.  To solve for $k$ let a partial derivative operator $D = \frac{\partial}{\partial k^T}$ operate on $||Xk-y||^2 = (Xk-y)^T(Xk-y) = k^T X^T X k - k^T X^T y - y^T X k + y^T y$ so that $D ||Xk-y||^2 = 2 X^T X k - 2 X^T Y = 0$ and $ k = (X^T X)^{-1} X^T Y$.  This formulation is termed whitened STA.

\biblio

\subsection*{Third order cardinal B-Splines}

%\subsection*{Splines coefficients parameterize kernels}
Central to STD is the ability to tune parametrized kernels to decrease the spike distance.  These kernels are spline functions composed of third order Cardinal B-splines.  This causes them to be everywhere differentiable with respect to time.  Starting with the recurrence relation in equation \eqref{eq:deBoor_full} (page 90 \cite{deboor}, page 143 \cite{farin2002curves}), the B-Spline coefficients from \cite{milovanovic2010calculation} are shown in equation \eqref{eq:spline}.  The number of steps within each knot sequence was set to $4$ time units.  The kernel is constructed by summing the multiple of all splines with their corresponding scaling parameters \eqref{eq:kernel}.  The $n$ order kernels $K_n$ are formed by B-splines $B_{i,n}$ which are the n-ary Cartesian product of one dimensional splines.  This is a deviation from the standard notation where $n$ refers instead to the order of the spline which was always $3$ in our experiments.

\begin{equation}
\label{eq:spline}
B_{0,3} =
\begin{cases}
 x \in [0,1) & \frac{1}{2}x^2
 \\ x \in [1,2) & -x^2 + 3x - \frac{3}{2}
 \\ x \in [2,3) & \frac{1}{2}x^2 - 3x + \frac{9}{2}
 \\ x \notin [0,3) & 0
\end{cases}
\end{equation}

\begin{equation} \label{eq:kernel}
K_n = \sum_{i=0} B_{i,n} \beta_i
\end{equation}

\biblio

Verifying the coefficients from \cite{milovanovic2010calculation} using the recursion formulas \eqref{eq:deBoor_recursion} and \eqref{eq:deBoor_ratio} from \cite{deboor} combine into \eqref{eq:deBoor_full}.  The final representation, displayed similarly to \cite{marschner2015fundamentals}, represents what the value of the spline is for each section of its support.  This is also commonly displayed as a matrix where elements represent polynomial coefficients \cite{milovanovic2010calculation}.  Setting $k=3$ leads to it being a third order and the knot sequence increasing as counting numbers makes these third order cardinal splines.  Successive support regions increment $j$ by $1$ causing an overlap of the nonzero region with the adjacent splines.  Adding additional evaluations within each knot range have the effect of smoothing the spline at the cost of taking up more time units.

\begin{equation} \label{eq:deBoor_recursion}
B_{j,k} := w_{j,k} B_{j,k-1} + (1 - w_{j+1,k})B_{j+1,k-1}
\end{equation}

\begin{equation} \label{eq:deBoor_ratio}
w_{j,k}(x) := \frac{x - t_j}{t_{j+k-1} - t_j}
\end{equation}

\begin{equation} \label{eq:deBoor_full}
B_{j,k} := \frac{x - t_j}{t_{j+k-1} - t_j}B_{j,k-1}
  + \frac{t_{j+k}-x}{t_{j+k} - t_{j+1}}B_{j+1,k-1}
\end{equation}

$$k=3$$

\begin{equation}
B_{j,3} := \frac{x - t_j}{t_{j+2} - t_j}B_{j,2}
  + \frac{t_{j+3}-x}{t_{j+3} - t_{j+1}}B_{j+1,2}
\end{equation}

\begin{equation}
B_{j,2} := \frac{x - t_j}{t_{j+1} - t_j}B_{j,1}
  + \frac{t_{j+2}-x}{t_{j+2} - t_{j+1}}B_{j+1,1}
\end{equation}

\begin{equation}
B_{j,1} := 
\begin{cases}
x \in [t_j, t_{j+1}) & 1 
\\ x \notin [t_j, t_{j+1}) & 0
\end{cases}
\end{equation}

\begin{equation}
j = 0
\hspace*{1cm}
t = [0,1,2,3]
\end{equation}

\begin{equation}
\begin{split}
B_{0,3} := \frac{x - t_0}{t_{2} - t_0}  B_{0,2}
  &+ \frac{t_{3}-x}{t_{3} - t_{1}}B_{1,2}
\\= \frac{x - 0}{2 - 0}  B_{0,2}
  &+ \frac{3-x}{3 - 1}B_{1,2}
\\= \frac{ x }{ 2 } B_{0,2}
  &+ \frac{ 3 - x }{ 2 } B_{1,2}
\end{split}
\end{equation}

\begin{equation}
\begin{split}
B_{0,2} := \frac{x - 0}{1 - 0}B_{0,1}
  +& \frac{2 - x}{2 - 1}B_{1,1}
\\= \frac{ x }{ 1 } B_{0,1}
  +& \frac{2 - x}{1}B_{1,1}
\end{split}
\end{equation}

\begin{equation}
B_{0,1} := 
\begin{cases}
x \in [0,1) & 1
\\x \notin [0,1) & 0
\end{cases}
\end{equation}

\begin{equation}
\begin{split}
B_{1,2} :=& \frac{ x - 1 }{ 2 - 1 } B_{1,1}
  + \frac{ 3 - x }{3 - 2 } B_{2,1}
\\=& \frac{ x - 1 }{ 1 } B_{1,0}
  + \frac{ 3 - x }{ 1 } B_{2,0}
\end{split}
\end{equation}

\begin{equation}
\begin{split}
B_{0,3} =& \frac{ x }{ 2 }
  (\frac{ x }{ 1 } B_{0,1} + \frac{2 - x}{1}B_{1,1})
\\ &+ \frac{ 3 - x }{ 2 }
  ( \frac{ x - 1 }{ 1 } B_{1,1} + \frac{ 3 - x }{ 1 } B_{2,1})
\end{split}
\end{equation}

\begin{equation}
B_{0,3} =
\begin{cases}
 x \in [0,1) & \frac{ x }{ 2 } \frac{ x }{ 1 }
 \\ x \in [1,2) &  \frac{ x }{ 2 } \frac{2 - x}{1}
      + \frac{ 3 - x }{ 2 } \frac{ x - 1 }{ 1 }
 \\ x \in [2,3) &  \frac{ 3 - x }{ 2 } \frac{ 3 - x }{ 1 }
 \\ x \notin [0,3) & 0
\end{cases}
\end{equation}

\begin{equation}
B_{0,3} =
\begin{cases}
 x \in [0,1) & \frac{1}{2}x^2
 \\ x \in [1,2) & -x^2 + 3x - \frac{3}{2}
 \\ x \in [2,3) & \frac{1}{2}x^2 - 3x + \frac{9}{2}
 \\ x \notin [0,3) & 0
\end{cases}
\end{equation}

\biblio

\subsection*{Spike trains as a vector space}\label{appendix:vs}

Spike trains can be generalized and transformed into a vector space.  This is achieved by augmenting them to include coefficients for each spike time along with defining appropriate addition and multiplication operations such that vector space properties are satisfied.  Sequence spaces are similar constructions that have a simpler addition operator due to their natural ordering.  These augmented spike trains are sequences of 2-tuples which have 5 primary properties: \eqref{eq:ss_1_r0}, \eqref{eq:ss_2_r0_t}, \eqref{eq:ss_3_tord}, \eqref{eq:ss_4_iN}, and \eqref{eq:ss_5_bounded}.  The times are exclusively in the past \eqref{eq:ss_2_r0_t} and are ordered so that they are strictly increasing \eqref{eq:ss_3_tord}.  The coefficients can be any non zero value in $\mathbb{R}$ \eqref{eq:ss_1_r0}.  These sequences can be finite or countably infinite \eqref{eq:ss_4_iN} and are confined to a subspace where the sum of the coefficients converges absolutely \eqref{eq:ss_5_bounded}.

\begin{equation}\label{eq:ss_1_r0}
\alpha_i \in \mathbb{R}  - \{0\}
\end{equation}
\begin{equation}\label{eq:ss_2_r0_t}
t_i \in \mathbb R_{> 0}
\end{equation}
\begin{equation}\label{eq:ss_3_tord}
t_i < t_j \forall i < j
\end{equation}
\begin{equation}\label{eq:ss_4_iN}
 i,j \in \mathbb{N}
\end{equation}
\begin{equation}\label{eq:ss_5_bounded}
\sum_{i=0}^\infty |\alpha_i| < \infty
\end{equation}

\textit{Definition}: Let $V$ be the space of finite or countably infinite sequences of time ordered 2-tuples $\bm{t} = \{(t_i,\alpha_i)\} \in V$ such that \eqref{eq:ss_1_r0}, \eqref{eq:ss_2_r0_t}, \eqref{eq:ss_3_tord}, \eqref{eq:ss_4_iN}, and \eqref{eq:ss_5_bounded} hold.

Adding these augmented spike trains results in a new sequence that consists of adding temporally corresponding coefficients and otherwise concatenating.  Defining two intermediary operators \eqref{eq:ss_a-b} and \eqref{eq:ss_cap} allows for a simpler addition definition for these sequences \eqref{eq:ss_add}.  The spike train vectors $\bm{t^A},\bm{t^B} \in V$  with times $t_i^A,t_j^B$ and coefficients $\alpha_i^A,\alpha_j^B$ correspond to different spike trains $A,B$.  Selecting tuples where the time only occurs in the first spike train can be expressed by a modified version \eqref{eq:ss_a-b} of the set difference operator ($A \setminus B$).  The set theory intersection operator $\cap$ can be redefined \eqref{eq:ss_cap} to select where the same time exists in both while adding coefficients and deleting tuples with zero coefficients.  The unioned sets in \eqref{eq:ss_add} are always disjoint so $\cup$ does not need to be modified.

\begin{equation}\label{eq:ss_a-b}
\bm{t^A} \setminus \bm{t^B} = 
\{ (t_i^A, \alpha_i^A) : 
(t_i^A, \alpha_i^A) \in \bm{t^A}
\text{ and } 
(t_j^B, \cdot ) \notin \bm{t^B}
\text{ and } 
t_i^A = t_j^B \}
\end{equation}
\begin{equation}\label{eq:ss_cap}
\bm{t^A} \cap \bm{t^B} = 
\{ (t_i^A, \alpha_i^A + \alpha_j^B) : 
(t_i^A, \alpha_i^A) \in \bm{t^A}
\text{ and } 
(t_j^B, \alpha_j^B) \in \bm{t^B}, t_i^A = t_j^B
\text{ and } 
\alpha_i^A + \alpha_j^B \neq 0 \}
\end{equation}
\begin{equation}\label{eq:ss_add}
\bm{t^A} + \bm{t^B} = 
\{ (t_i, \alpha_i) \in 
\{\bm{t^A} \setminus \bm{t^B}\} \cup 
\{\bm{t^B} \setminus \bm{t^A}\} \cup
\{\bm{t^A} \cap \bm{t^B}\} 
\}
\end{equation}
\begin{equation}\label{eq:ss_sm}
a \bm{t^A} = \{ (t_i^A, a \alpha_i^A) : (t_i^A, \alpha_i^A) \in \bm{t^A},  a  \alpha_i^A \neq 0 \}
\end{equation}

Scalar multiplication with a vector \eqref{eq:ss_sm}, is only applied to the spike's coefficient and not the time.  A doubling would then correspond to a sequence of spikes at the same times but with each coefficient being twice as large.  Setting $a = 0$ removes all elements. We now show that for any $\bm{t^A},\bm{t^B} \in V$ the operations $\bm{t^A}+\bm{t^B}$ and $a \bm{t^A}$ satisfy all of the properties \eqref{eq:ss_1_r0}...\eqref{eq:ss_5_bounded} and thus the space is closed under vector addition and scalar multiplication.

%\subsection*{Closure}
%\subsubsection*{Closure under vector addition}
Adding spike trains does not alter the set to which the coefficients belong \eqref{eq:ss_1_r0} since zeros are deleted by construction.  Times are unaffected by \eqref{eq:ss_add} except when removed, thus \eqref{eq:ss_2_r0_t} and \eqref{eq:ss_3_tord}.  The indices of spikes can change but are still countable \eqref{eq:ss_4_iN}.  The spike train resulting from addition is just the concatenation of all spike trains $\bm{t^A} \setminus \bm{t^B}$, $\bm{t^B} \setminus \bm{t^A}$, and $\bm{t^A} \cap \bm{t^B}$ with non zero coefficients $\alpha_i^A + 0$, $0 + \alpha_i^B$, and $\alpha_i^A + \alpha_i^B$ respectively.  Thus \eqref{eq:ss_5_bounded} is now a trivial application of Minkowski's inequality where $p=1$: $\sum | \alpha_i^A + \alpha_j^B | \leq \sum | \alpha_i^A | + \sum| \alpha_j^B | < \infty$ (page 103 \cite{meise1997introduction}, page 588 \cite{schechter1996handbook}).

%\subsubsection*{Closure under scalar multiplication}
Scalar multiplication results in a new vector with tuple elements in the same set \eqref{eq:ss_1_r0} \eqref{eq:ss_2_r0_t}.  As before with addition, zero elements are explicitly deleted so that multiplying a vector by zero has the effect of removing all spikes.  The scalar has no effect on times, except when it is zero and that effect is to remove spikes so the temporal ordering is maintained \eqref{eq:ss_3_tord}.  The indices are untouched, except when all spikes are removed, in either case \eqref{eq:ss_4_iN} holds.  In \eqref{eq:ss_5_bounded}, the scalar simply factors out of the summation yielding $|a| \sum_{i=0}^\infty |\alpha_i^A| < \infty$ which is still clearly a finite sum.

%\subsubsection*{Vector space properties}
A vector space is made of a field ($\mathbb{R}$), a set of vector objects ($V$), and the following rules: commutativity, associativity, identities, inverse, compatibility, and distributivity \cite{hoffmanlinear}.  The operations we describe satisfy all of these properties, therefore $V$ is a vector space.

\begin{itemize}
\item Commutativity: $\bm{t^A} + \bm{t^B} = \bm{t^B} + \bm{t^A}$.  \eqref{eq:ss_add} is symmetric by construction and commutes.
\item Associativity: $\bm{t^A} + (\bm{t^B} + \bm{t^C}) = (\bm{t^A} + \bm{t^B}) + \bm{t^C}$. Both of these expressions describe the same sequence due to their set theory expansion.
\item Additive Identity: $\bm{t^A} + \bm{0} = \bm{t^A}$.  The zero vector is simply the empty sequence.
\item Additive Inverse: This occurs when $a = -1$ in $\bm{t^A} + a \bm{t^A} = \bm{0}$ leading to all terms canceling and producing the empty vector $\{(t_i^A, \alpha_i^A + -\alpha_i^A)\} = \{\} = \bm{0}$.
\item Multiplicative Identity: $1\bm{t^A} = \bm{t^A}$, where $1$ is the multiplicative identity of the field $\mathbb{R}$ and $\{(t_i^A, 1 \alpha_i^A)\} = \{(t_i^A, \alpha_i^A)\}$.
\item Compatibility:  $a(b\bm{t^A}) = (ab)\bm{t^A}$ for $a,b \in \mathbb{R}$ and $\bm{t^A} \in V$ yields $\{(t_i^A, a(b \alpha_i^A))\} = \{(t_i^A, (ab)\alpha_i^A)\}$, which reduces to field scalar compatibility.
\item Distributivity (scalar addition): $(a + b)\bm{t^A} = a\bm{t^A} + b\bm{t^A}$.  $\{(t_i^A, (a+b)\alpha_i^A\} = \{(t_i^A, a\alpha_i^A+b\alpha_i^A\} = \{(t_i^A, a\alpha_i^A)\} + \{(t_i^A, b\alpha_i^A)\}$
\item Distributivity (vector addition): $a(\bm{t^A} + \bm{t^B}) = a\bm{t^A} + a\bm{t^B}$.  The left side is a scalar multiple applied to the three separate partitions in \eqref{eq:ss_add}.  The term $a\bm{t^A}$ is then the spikes with corresponding $\alpha_i^A$ components from both $\bm{t^A} \setminus \bm{t^B}$ and $\bm{t^A} \cap\bm{t^B}$ scaled by $a$. Likewise for $a\bm{t^B}$.
\end{itemize}

\biblio

\subsection*{Spike trains as an inner product space}

The inspiration for this work originates from a similar technique \cite{banerjee2016learning} which uses a measure of disparity between spike trains.  While other such measures exist, this measure was selected for its ability to model the vanishing and asymmetric aspect of past incoming spikes along with the feasibility of obtaining a closed form gradient.  This starts by selecting an appropriately expressive function which can capture enough of the dynamics while also simplifying the analysis.  The goal here is to show that an emerging inner product induces a metric in which the subset of finite spike trains is embedded.

This disparity measure leads to a gradient with respect to the parameters which is core to both techniques.  The primary difference being that this work learns the parameters of a response function instead of synaptic weights.  This results in different applications.  A parameterized expression in terms of time models the shape along with the asymmetric and vanishing aspects of post synaptic potentials \eqref{eq:nc4.1}.  Integrating over all parameters for two spikes gives an expression in terms of spike times \eqref{eq:nc4.4}.

\begin{equation} \label{eq:nc4.1}
f_{\beta, \tau} (t) = \frac{1}{\tau} e^\frac{-\beta}{t} e^\frac{-t}{\tau}
\hspace*{.5cm}
\text{  for  }
\hspace*{.5cm}
\beta, \tau \ge 0
\hspace*{.5cm}
\text{  and  }
\hspace*{.5cm}
t > \epsilon > 0
\hspace*{1cm}
\text{4.1 from \cite{banerjee2016learning}}
\end{equation}

\begin{equation} \label{eq:nc4.4}
\int_0^\tau \int_0^\infty 
\frac{1}{\tau} e^\frac{-\beta}{t_1} e^\frac{-t_1}{\tau}
\times
\frac{1}{\tau} e^\frac{-\beta}{t_2} e^\frac{-t_2}{\tau}
d\beta d\tau
= 
\frac{t_1 \times t_2}{(t_1 + t_2)^2} 
e^{-\frac{t_1 + t_2}{\tau}}
\hspace*{1cm}
\text{4.4 from \cite{banerjee2016learning}}
\end{equation}

The following properties needed to make an inner product are symmetry, linearity, and positive definiteness. It will be shown that \eqref{eq:IP_int} satisfies these properties and is therefore an inner product.  The reordering of the integrals and the summations from \eqref{eq:IP_int} to \eqref{eq:nc4.3_fub} follows from Fubini's theorem since \eqref{eq:IP} will be shown to be less than infinity.  The alphas are a multiple of two absolutely converging sequences \eqref{eq:ss_5_bounded} is thus bounded (by Theorem 3.50 \cite{rudin1964principles}).  Times are positive so terms $\frac{t_i^A \times t_j^B}{(t_i^A + t_j^B)^2}$ and $e^{-\frac{t_i^A + t_j^B}{\tau}}$ are bounded below by zero and above by one.  The inclusion of each term has the effect of shrinking the sum and so equation \eqref{eq:IP} converges.  This validates the use of Fubini's theorem and the simplification from \eqref{eq:IP_int} to \eqref{eq:nc4.3_fub}.

\begin{equation} \label{eq:IP_int}
\langle \bm{t^A}, \bm{t^B} \rangle 
 = 
\int_0^\tau \int_0^\infty \left( \sum_{i=1}^\infty \alpha_i f_{\beta, \tau} (t_i^A) \right) \left( \sum_{i=1}^\infty \alpha_i f_{\beta, \tau} (t_i^B) \right) d\beta d\tau
\end{equation}

\begin{equation} \label{eq:nc4.3_fub}
\langle \bm{t^A}, \bm{t^B} \rangle 
 = 
\sum_{i,j=1}^{\infty}
\alpha_i \times \alpha_j
\int_0^\tau \int_0^\infty 
f_{\beta, \tau} (t_i^A) 
\times
f_{\beta, \tau} (t_j^B)
d\beta d\tau
\end{equation}

\begin{equation} \tag{\ref{eq:IP}}
\langle \bm{t^A}, \bm{t^B} \rangle = 
\sum_{i,j=1}^{\infty} (\alpha_i^A \times \alpha_j^B) 
\frac{t_i^A \times t_j^B}{(t_i^A + t_j^B)^2} 
e^{-\frac{t_i^A + t_j^B}{\tau}}
\end{equation}

It is symmetric because switching the parameters yields an equivalent result since the nested operations are all individually symmetric.  The scalar multiple factors out of the summation, thus: $\langle c \bm{t^A}, \bm{t^B} \rangle = c \langle \bm{t^A}, \bm{t^B} \rangle$.  For additivity, let the vectors be altered to include virtual zeros where there's a corresponding non zero $\alpha$ in the other.  These zeros have no effect except to allow for clearer separation.  At each point in the summation $\alpha_i^A + \alpha_i^C$ can be split into two separate double summations such that $\langle \bm{t^A} + \bm{t^C}, \bm{t^B} \rangle = \langle \bm{t^A}, \bm{t^B} \rangle + \langle \bm{t^C}, \bm{t^B} \rangle$.

If $\bm{t^A} \neq \bm{0}$ and $\langle \bm{t^A}, \bm{t^A} \rangle > 0$ then it is positive definite.  Equation \eqref{eq:IP_int} integrates the square of a nonlinear function $(F_{\beta,\tau}(\bm{t}))^2 = (\sum_i \alpha_i f_{\beta,\tau}(t_i))^2 \ge 0$.  In order for it to be zero everywhere it would have to be the zero function, but clearly it is not.   Therefore, it is greater than zero at least somewhere causing \eqref{eq:IP} to be an inner product on the vector space of augmented spike trains.  The usual spike trains are simply the subset of finite sequences with the coefficients set to $1$ and ignored.  These need to be finite since the countable sequence is unbounded and not in the vector space.  This subset is a metric space with the same metric \cite{rudin1964principles} \eqref{eq:distance} where $E$ an error between spike trains to be minimized. 

\begin{equation} \tag{\ref{eq:distance}}
E = d^2 = \langle \bm{t^A} - \bm{t^B}, \bm{t^A} - \bm{t^B} \rangle
\end{equation}

\biblio

\subsection*{Gradient calculation}

Inspired by a similar technique\cite{banerjee2016learning}, the full Volterra series operator for the GCSRM is perturbed and set equal to the first order Taylor approximation \eqref{eq:GCSRMp}.  A simplifying step \eqref{eq:GCSRMc} cancels terms.  The terms where two perturbation terms appear are additionally removed.  Solving for $\Delta t_l^O$ \eqref{eq:GCSRMdt} reveals how the spike time changes with respect to changes in the past system.  After detailing this general formulation, to clarify further, the simpler first order kernel derivation is shown.  Starting again from the threshold equation, we proceed through the partial derivatives until the gradient update calculations.

\begin{equation} \label{eq:GCSRMp}
\begin{split}
\tilde{\Theta} = & K_0 + \sum_{n=1}^N \idotsint_n K(\tau_{1 \rightarrow n}; \beta_{i,l}) \prod_{i = 1}^n x(t_l^O - \tau_i) d\tau_i + \sum \eta(t_l^O - t_k^O; \mu_k)
\\ \\= &(K_0 + \Delta K_0)+ \sum_{n=1}^N \idotsint_n \Big[K(\tau_{1 \rightarrow n}; \beta_{i,l} + \Delta \beta_{i,l})\Big] 
\prod_{j = 1}^n
\Big[x(t_l^O + \Delta t_l^O - \tau_j)
\Big] 
d\tau_j
\\ &
+ \sum \eta(t_l^O + \Delta t_l^O - t_k^O - \Delta t_k^O; \mu_k + \Delta \mu_k)
\\ \\ = & (K_0 + \Delta K_0)
\\ & + \sum_{n=1}^N  \idotsint_n
\Big[ K(\tau_{1 \rightarrow n}; \beta_{i,l}) + \sum B_i(\tau_{1 \rightarrow n}) \Delta \beta_{i,l}\Big]
\prod_{j = 1}^n
\Big[ x(t_l^O - \tau_j) + \frac{\partial x}{\partial t} \Big|_{t_l^O - \tau_j} \Delta t_l^O 
\Big] 
d\tau_j
\\ &
+ \sum \Big[ \eta(t_l^O - t_k^O; \mu_k)
+ \frac{\partial \eta}{\partial t} \Big|_{t_l^O - t_k^O} (\Delta t_l^O - \Delta t_k^O) 
+ \frac{\partial \eta}{\partial \mu} \Big|_{t_l^O - t_k^O} (\Delta \mu_k) \Big]
\end{split}
\end{equation}

\begin{equation} \label{eq:GCSRMc}
\begin{split}
0 = \Delta K_0 &
+ 
\sum_{n=1}^N  \idotsint_n
\Big[\sum B_i(\tau_{1 \rightarrow n}) \Delta \beta_{i,l}\Big]
\Big[
\prod_{j = 1}^n
x(t_l^O - \tau_j)
d\tau_j
\Big]
\\ & + 
\sum_{n=1}^N  \idotsint_n
K(\tau_{1 \rightarrow n}; \beta_{i,l})
\Big[
\sum_{m=1}^n
\Big[ 
\prod_{j \ne m}^n
x(t_l^O - \tau_j)
\Big]
\frac{\partial x}{\partial t} \Big|_{t_l^O - \tau_m} \Delta t_l^O
\Big]
\prod_{j = 1}^n
d\tau_j
\\ &
+ \sum \Big[ \frac{\partial \eta}{\partial t} \Big|_{t_l^O - t_k^O} (\Delta t_l^O - \Delta t_k^O) 
+ \frac{\partial \eta}{\partial \mu} \Big|_{t_l^O - t_k^O} (\Delta \mu_k)\Big]
\end{split}
\end{equation}

\begin{equation} \label{eq:GCSRMdt}
\resizebox{.9\hsize}{!}{$
\Delta t_l^O = 
\frac{
\sum \frac{\partial \eta}{\partial t} \big|_{t_l^O - t_k^O}  \Delta t_k^O
-
\sum \frac{\partial \eta}{\partial \mu} \big|_{t_l^O - t_k^O} \Delta \mu_k
-
\Delta K_0
- 
\sum_{n=1}^N  \idotsint_n
(\sum B_i(\tau_{1 \rightarrow n}) \Delta \beta_{i,l})
\prod_{j = 1}^n
x(t_l^O - \tau_j)
d\tau_j
}
{
\sum_{n=1}^N  \idotsint_n
K(\tau_{1 \rightarrow n}; \beta_{i,l})
\Big[
\sum_{m=1}^n
\Big[ 
\prod_{j \ne m}^n
x(t_l^O - \tau_j)
\Big]
\frac{\partial x}{\partial t} \big|_{t_l^O - \tau_m} 
\Big]
\prod_{j = 1}^n
d\tau_j
+ \sum \frac{\partial \eta}{\partial t} \big|_{t_l^O - t_k^O}
}
$}
\end{equation}

Selecting only the first order kernel ($N=1$) and starting back at the beginning \eqref{eq:core} we show the core derivation used in these experiments.  As before, the parameters are perturbed.

\begin{equation} \label{eq:core}
\tilde{\Theta} = \int K(\tau; \beta_{i,l}) x(t_l^O - \tau) d\tau + \sum \eta(t_l^O - t_k^O; \mu_k)
\end{equation}

\begin{equation} \label{eq:perturbed}
\begin{split}
= & \int K(\tau; \beta_{i,l} + \Delta \beta_{i,l}) x(t_l^O + \Delta t_l^O - \tau) d\tau 
\\ &
+ \sum \eta(t_l^O + \Delta t_l^O - t_k^O - \Delta t_k^O; \mu_k + \Delta \mu_k)
\end{split}
\end{equation}

A first order taylor approximation is performed for each of the perturbed variables which emerges from the definition of a derivative:
\begin{equation}
f'(x) = lim_{\Delta x \rightarrow 0} \frac{f(x + \Delta x) - f(x)}{\Delta x} \end{equation}
At the limit this resolves to $f(x + \Delta x) = f(x) + \Delta x f'(x)$.  Substituting this for our variables yields equation \eqref{eq:FOTA} where the notation $\Big|_{t}$ represents the location where the function is evaluated.

\begin{equation} \label{eq:FOTA}
\begin{split}
= & \int (K(\tau; \beta_{i,l}) + \sum B_i(\tau) \Delta \beta_{i,l}) (x(t_l^O - \tau) + \frac{\partial x}{\partial t} \Big|_{t_l^O - \tau} \Delta t_l^O) d\tau 
\\ &
+ \sum (\eta(t_l^O - t_k^O; \mu_k)
+ \frac{\partial \eta}{\partial t} \Big|_{t_l^O - t_k^O} (\Delta t_l^O - \Delta t_k^O) 
+ \frac{\partial \eta}{\partial \mu} \Big|_{t_l^O - t_k^O} (\Delta \mu_k) ) 
\end{split}
\end{equation}

%%%%%%%%%%%%%%%%%%%%%%%%%%%%%%
%\rule{13cm}{0.4pt}

Setting \eqref{eq:core} equal to \eqref{eq:FOTA} and canceling out terms yields \eqref{eq:FOTA_w_core}.

\begin{equation} \label{eq:FOTA_w_core}
\begin{split}
0 = & \int \Big(\sum B_i(\tau) \Delta \beta_{i,l}\Big)x(t_l^O - \tau)d\tau 
\\ &
+ \int K(\tau; \beta_{i,l}) \frac{\partial x}{\partial t} \Big|_{t_l^O - \tau} \Delta t_l^O d\tau 
\\ &
+ \sum \frac{\partial \eta}{\partial t} \Big|_{t_l^O - t_k^O} (\Delta t_l^O - \Delta t_k^O) 
+ \sum \frac{\partial \eta}{\partial \mu} \Big|_{t_l^O - t_k^O} \Delta \mu_k
\end{split}
\end{equation}

%%%%%%%%%%%%%%%%%%%%%%%%%%%%%%
%\rule{15cm}{0.4pt}
An intermediary step of isolating $\Delta t_l^O$ is shown in \eqref{eq:FOTA_solving}.  This can be factored out and becomes the denominator in \eqref{eq:dtl}.

\begin{equation} \label{eq:FOTA_solving}
\begin{split}
\int K(\tau; \beta_{i,l}) \frac{\partial x}{\partial t} \Big|_{t_l^O - \tau} & \Delta  t_l^O d\tau 
\\  + \sum \frac{\partial \eta}{\partial t} \Big|_{t_l^O - t_k^O} & \Delta t_l^O 
\end{split}
\hspace*{.5cm}
= 
\hspace*{.5cm}
\begin{split}
& \sum \frac{\partial \eta}{\partial t} \Big|_{t_l^O - t_k^O} \Delta t_k^O 
\\ & - \sum \frac{\partial \eta}{\partial \mu} \Big|_{t_l^O - t_k^O} \Delta \mu_k
\\ & - \int \Big(\sum B_i(\tau) \Delta \beta_{i,l}\Big)x(t_l^O - \tau)d\tau 
\end{split}
\end{equation}

%%%%%%%%%%%%%%%%%%%%%%%%%%%%%%
%\rule{13cm}{0.4pt}

Now that an expression for $\Delta t_l^O$ has been obtained, partial derivatives with respect to the parameters can be taken.  The partial derivatives in \eqref{eq:dtdb}, \eqref{eq:dtdu}, and \eqref{eq:dtdt} represent how the timing of spike $t_l^O$ changes with respect to $\beta$, $\mu$, and previous spikes $t_k^O$ respectively.  The limits of integration are from the current moment to the end of the kernel.

\begin{equation} \label{eq:dtl}
\Delta t_l^O 
= 
\frac{
\sum \frac{\partial \eta}{\partial t} \Big|_{t_l^O - t_k^O} \Delta t_k^O 
- \sum \frac{\partial \eta}{\partial \mu} \Big|_{t_l^O - t_k^O} \Delta \mu_k
- \int \Big(\sum B_i(\tau) \Delta \beta_{i,l}\Big)x(t_l^O - \tau)d\tau 
}{
\int K(\tau; \beta_{i,l}) \frac{\partial x}{\partial t} \Big|_{t_l^O - \tau} d\tau 
+ \sum \frac{\partial \eta}{\partial t} \Big|_{t_l^O - t_k^O}
}
\end{equation}

\begin{equation} \tag{\ref{eq:dtdb}}
\frac{\partial t_l^O}{\partial \beta_{i,l}} 
= 
\frac{
- \int B_i(\tau) x(t_l^O - \tau) d\tau 
}{
\int_0^{|K|} K(\tau; \beta_{i,l}) \frac{\partial x}{\partial t} \Big|_{t_l^O - \tau} d\tau 
+ \sum \frac{\partial \eta}{\partial t} \Big|_{t_l^O - t_k^O}
}
\end{equation}

\begin{equation} \tag{\ref{eq:dtdu}}
\frac{\partial t_l^O}{\partial \mu_l} 
= 
\frac{
- \frac{\partial \eta}{\partial \mu} \Big|_{t_l^O - t_k^O}
}{
\int_0^{|K|} K(\tau; \beta_{i,l}) \frac{\partial x}{\partial t} \Big|_{t_l^O - \tau} d\tau 
+ \sum \frac{\partial \eta}{\partial t} \Big|_{t_l^O - t_k^O}
}
\end{equation}

\begin{equation} \tag{\ref{eq:dtdt}}
\frac{\partial t_l^O}{\partial t_k^O} 
= 
\frac{
\frac{\partial \eta}{\partial t} \Big|_{t_l^O - t_k^O} }{
\int_0^{|K|} K(\tau; \beta_{i,l}) \frac{\partial x}{\partial t} \Big|_{t_l^O - \tau} d\tau 
+ \sum \frac{\partial \eta}{\partial t} \Big|_{t_l^O - t_k^O}
}
\end{equation}

%%%%%%%%%%%%%%%%%%%%%%%%%%%%%%
%\rule{13cm}{0.4pt}

The following equations \eqref{eq:Dtdb}, \eqref{eq:Dtdu}, \eqref{eq:dEdb}, and \eqref{eq:dEdu} connect the partial derivatives through the chain rule to the error via \eqref{eq:dEdt} \cite{banerjee2016learning}.  The definitions are self referencing and recursive for \eqref{eq:Dtdb} and \eqref{eq:Dtdu} where the first instance is just the corresponding partial from before.  The expressions \eqref{eq:dEdb} and \eqref{eq:dEdu} represent how the spike train distance changes with respect to the $\beta$ and $\mu$ parameters respectively.

\begin{equation} \tag{\ref{eq:Dtdb}}
\frac{D t_{k+1}^O}{\partial \beta_{i,l}} 
= 
\sum_{t_j > t_l}^{k}
\frac{D t_j^O}{\partial \beta_{i,l}}
\frac{\partial t_{k+1}^O}{\partial t_j^O}
\end{equation}

\begin{equation} \label{eq:Dtdu}
\frac{D t_{p+1}^O}{\partial \mu_{k}} 
= 
\sum_{t_j > t_p}^{p}
\frac{D t_j^O}{\partial \mu_k}
\frac{\partial t_{p+1}^O}{\partial t_j^O}
\end{equation}

\begin{equation} \label{eq:dEdt}
\begin{split}
\frac{\partial E}{\partial t_i^O} = 
  2 & \Bigg(
\sum_{j=1}^N 
\frac{ 
	t_j^O((t_j^O-t_i^O) - \frac{t_i^O}{\tau} (t_j^O + t_i^O) 
	}{ 
	(t_j^O + t_i^O)^3 
	}
	e^{-\frac{t_j^O+t_i^O}{\tau}} 
\\ & -
\sum_{j=1}^N 
\frac{ 
	t_j^D((t_j^D-t_i^O) - \frac{t_i^O}{\tau} (t_j^D + t_i^O) 
	}{ 
	(t_j^D + t_i^O)^3 
	} 
	e^{-\frac{t_j^D+t_i^O}{\tau}} 
\Bigg)
\end{split}
\end{equation}

\begin{equation} \label{eq:dEdb}
\frac{\partial E}{\partial \beta_{i,l}} 
= \sum_{k \in F}
\frac{\partial E}{\partial t_{k}^O}
\frac{D t_{k}^O}{\partial \beta_{i,l}} 
\end{equation}

\begin{equation} \label{eq:dEdu}
\frac{\partial E}{\partial \mu_k} 
= \sum_{k \in F}
\frac{\partial E}{\partial t_{k}^O}
\frac{D t_{k}^O}{\partial \mu_k} 
\end{equation}

%%%%%%%%%%%%%%%%%%%%%%%%%%%%%%
%\rule{13cm}{0.4pt}

The equations \eqref{eq:db} and \eqref{eq:du} are the vanilla gradient descent update rules.  The terms $\alpha_\beta$ and $\alpha_\mu$ are the respective learning rates while the summation represents the gradient.

\begin{equation} \tag{\ref{eq:db}}
\beta_i = \beta_i - \alpha_\beta \sum_{k \in F_l}
\frac{\partial E}{\partial \beta_{i,l}}
\end{equation}

\begin{equation} \tag{\ref{eq:du}}
\mu = \mu - \alpha_\mu \sum_{k \in F_l}
\frac{\partial E}{\partial \mu_k}
\end{equation}

In practice, vanilla gradient descent can be too slow.  One way to speed this up is by first updating a momentum term $p$ with respect to the variable $E$ \eqref{eq:momentum}.  The momentum then replaces the gradient in equations \eqref{eq:db} and \eqref{eq:du}.  When $\alpha_p = 0$ this reduces to vanilla gradient descent.

\begin{equation} \label{eq:momentum}
p_{i+1} = \alpha_p p_i + \nabla E
\end{equation}

\biblio

\subsection*{Second order gradient calculation}

The analysis for $2^{nd}$ order kernels is essentially identical, excluding simple differences in algebra.  The core changes are in the setup \eqref{eq:2nd_CSRM} and in denominator of \eqref{eq:dt_2d} which matches the partial derivatives.  The numerator for the partials are the same except for $\beta$.  The chain rule and update equations are the same.  Further generalizations continue to follow this pattern.

\begin{equation} \label{eq:2nd_CSRM}
\tilde{\Theta} = 
\int\int 
K(\tau; \beta_{i,l}) x(t_l^O - \tau_1) x(t_l^O - \tau_2) d\tau_1 d\tau_2 
+ \sum \eta(t_l^O - t_k^O; \mu_k)
\end{equation}

\begin{equation} \label{eq:dt_2d}
\resizebox{.9\hsize}{!}{$
\Delta t_l^O
= 
\frac{
\sum \frac{\partial \eta}{\partial t} \big|_{t_l^O - t_k^O} \Delta t_k^O
- \sum \frac{\partial \eta}{\partial \mu} \big|_{t_l^O - t_k^O} \Delta \mu_k
- \int \int
(\sum B_i(\tau_1, \tau_2) \Delta \beta_{i,l}) x(t_l^O - \tau_1) x(t_l^O - \tau_2)
d\tau_1 d\tau_2
}{
\int \int
K(\tau_1, \tau_2; \beta_{i,l}) x(t_l^O - \tau_1) \frac{\partial x}{\partial t} \big|_{t_l^O - \tau_2}
d\tau_1 d\tau_2
+\int \int
K(\tau_1, \tau_2; \beta_{i,l}) \frac{\partial x}{\partial t} \big|_{t_l^O - \tau_1} x(t_l^O - \tau_2)
d\tau_1 d\tau_2
+ \sum \frac{\partial \eta}{\partial t} \big|_{t_l^O - t_k^O}
}$}
\end{equation}

\biblio

%\subfile{./tex/appendix-chain_rule.tex}

\biblio

\iffalse
\fi

%https://en.wikibooks.org/wiki/LaTeX/Modular_Documents#Subfiles_and_bibtex
%\bibliographystyle{plain}
\bibliographystyle{naturemag}
\bibliography{./refs}

\end{document}